\documentclass[12pt,preprint]{emulateapj}

\begin{document}
\title{New Determinations of the UV Luminosity Functions from $z\sim 9$ to $z\sim 2$ show a remarkable consistency with halo growth and a constant star formation efficiency}
\author{R.J. Bouwens\altaffilmark{1}, P.A. Oesch\altaffilmark{2,3}, M. Stefanon\altaffilmark{1},
  G. Illingworth\altaffilmark{4}, I. Labb{\'e}\altaffilmark{5}, N. Reddy\altaffilmark{6}, H. Atek\altaffilmark{7}, 
  M. Montes\altaffilmark{8}, R. Naidu\altaffilmark{9},
  T. Nanayakkara\altaffilmark{5}, E. Nelson\altaffilmark{10}, S. Wilkins\altaffilmark{11}}
\altaffiltext{1}{Leiden Observatory, Leiden University, NL-2300 RA Leiden, Netherlands}
\altaffiltext{2}{Department of Astronomy, University of Geneva, Chemin Pegasi 51, 1290 Versoix, Switzerland}
\altaffiltext{3}{Cosmic Dawn Center (DAWN), Niels Bohr Institute, University of Copenhagen, Jagtvej 128, K\o benhavn N, DK-2200, Denmark}
\altaffiltext{4}{UCO/Lick Observatory, University of California, Santa Cruz, CA 95064}
\altaffiltext{5}{Centre for Astrophysics \& Supercomputing, Swinburne
  University of Technology, PO Box 218, Hawthorn, VIC 3112, Australia}
\altaffiltext{6}{University of California, Riverside, CA 92521, USA}
\altaffiltext{7}{Institut d'Astrophysics de Paris, 98bis Boulevard Arago, 75014 Paris, France}
\altaffiltext{8}{Space Telescope Science Institute, 3700 San Martin Drive, Baltimore, MD 21218}
\altaffiltext{9}{Center for Astrophsics, 60 Garden St, Cambridge, MA 02138, United States}
\altaffiltext{10}{Astrophysical \& Planetary Sciences, 391 UCB, 2000 Colorado Ave, Boulder, CO 80309, Duane Physics Building, Rm. E226}
\altaffiltext{11}{Department of Physics \& Astronomy, University of Sussex, Falmer, Brighton, BN1 9QH, United Kingdom}

\begin{abstract}
Here we provide the most comprehensive determinations of the
rest-frame $UV$ LF available to date with {\it HST} at $z\sim2$, 3, 4,
5, 6, 7, 8, and 9.  Essentially all of the non-cluster extragalactic
legacy fields are utilized, including the Hubble Ultra Deep Field
(HUDF), the Hubble Frontier Field parallel fields, and all five
CANDELS fields, for a total survey area of 1136 arcmin$^2$.  Our
determinations include galaxies at $z\sim2$-3 leveraging the deep
HDUV, UVUDF, and ERS WFC3/UVIS observations available over a $\sim$150
arcmin$^2$ area in the GOODS North and GOODS South regions.  All
together, our collective samples include $>$24,000 sources,
$>$$2.3\times$ larger than previous selections with {\it HST}.  5766,
6332, 7240, 3449, 1066, 601, 246, and 33 sources are identified at
$z\sim2$, 3, 4, 5, 6, 7, 8, and 9, respectively.  Combining our
results with an earlier $z\sim10$ LF determination by Oesch et
al.\ (2018a), we quantify the evolution of the $UV$ LF.  Our results
indicate that there is (1) a smooth flattening of the faint-end slope
$\alpha$ from $\alpha\sim-2.4$ at $z\sim10$ to $-1.5$ at $z\sim2$, (2)
minimal evolution in the characteristic luminosity $M^*$ at $z\geq
2.5$, and (3) a monotonic increase in the normalization $\log_{10}
\phi^*$ from $z\sim10$ to $z\sim2$, which can be well described by a
simple second-order polynomial, consistent with an ``accelerated''
evolution scenario.  We find that each of these trends (from $z\sim10$
to $z\sim2.5$ at least) can be readily explained on the basis of the
evolution of the halo mass function and a simple constant star
formation efficiency model.
\end{abstract}

\section{Introduction}

Quantifying the build-up of galaxies in the early universe remains one
of a principal area of interest in extragalactic astronomy involves
(e.g., Madau \& Dickinson 2014; Davidzon et al.\ 2017).  Studies of
galaxy build-up have become increasingly mature, with ever more
detailed efforts to measure the star formation rates and stellar
masses of galaxies (e.g., Salmon et al.\ 2015; Leja et al.\ 2019;
Stefanon et al.\ 2021, in prep).  Determinations of the volume density
in the context of star formation rate and stellar mass measurements
allow for connections to the underlying dark matter halos (e.g.,
Behroozi et al.\ 2013; Harikane et al.\ 2016, 2018; Stefanon et
al.\ 2017a).

One prominent, long-standing gauge of galaxy build-up is the
luminosity function of galaxies in the rest-frame $UV$, which
represents the volume density of galaxies as a function of the $UV$
luminosity.  As the time-averaged star formation rate of galaxies is
proportional to the unobscured luminosities of galaxies in the
rest-frame $UV$, the $UV$ luminosity function provides us with a
measure of how quickly galaxies grow with cosmic time.

There is already significant work on the $UV$ LF across a wide range
in redshifts, from local studies to studies in the early universe.
Broadly, the normalization $\phi^*$ and faint-end slope $\alpha$ of
the $UV$ LF have been found to increase and to flatten, respectively,
with cosmic time (Bouwens et al.\ 2015, 2017; Finkelstein et
al.\ 2015; Bowler et al.\ 2015; Parsa et al.\ 2016; Ishigaki et
al.\ 2018), while the characteristic luminosity remains fixed with
cosmic time (Bouwens et al.\ 2015, 2017; Finkelstein et al.\ 2015;
Bowler et al.\ 2015; Parsa et al.\ 2016) or becomes fainter (Arnouts
et al.\ 2005).  Motivated by many theoretical models, Bouwens et
al.\ (2015) showed that the evolution of the faint-end slope from
$z\sim8$ to $z\sim4$ could be naturally explained by a similar
steepening of the halo mass function over the relevant range (see also
Mason et al.\ 2015; Tacchella et al.\ 2013, 2018).

\begin{figure*}
\epsscale{1.15}
\plotone{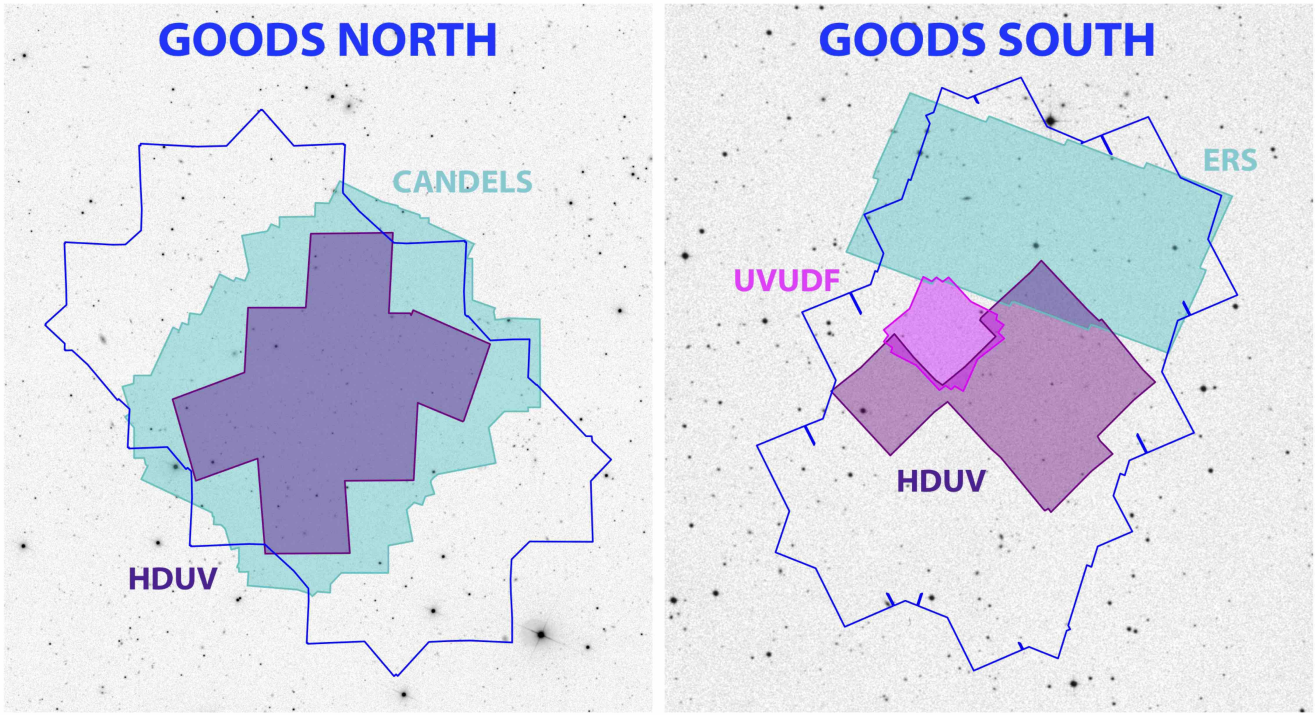}
\caption{The layout of the search fields we utilize with WFC3/UVIS
  $UV_{275}U_{336}$ $\sim$0.25-0.4$\mu$m data to identify $z\sim2$-3
  galaxies.  These include the $\sim$93 arcmin$^2$ HDUV fields (Oesch
  et al.\ 2018b), the $\sim$7 arcmin$^2$ UVUDF field (Teplitz et
  al.\ 2013), and the $\sim$50 arcmin$^2$ ERS (Windhorst et al.\ 2011)
  data set.  The cyan footprint shown over the GOODS-North shows the
  WFC3/UVIS imaging data available from the CANDELS program in the
  F275W (Grogin et al.\ 2011) and has an exposure time equivalent to
  $\sim$6 orbits (Oesch et al.\ 2018b).  All together we leverage a
  $\sim$150 arcmin$^2$ search area to identify $>$11,000 $z\sim2$-3
  galaxies over GOODS South and GOODS North.\label{fig:layout}}
\end{figure*}

Given the increasing clarity in the general evolutionary trends in the
$UV$ LF with redshift, galaxy evolution studies are entering an era
where precision measurements become increasingly key.  To date, there
has been no systematic, self-consistent determination of the evolution
of the rest-frame $UV$ LF from $z\sim9$ to $z\sim2$.

The availability of deep wide-area WFC3/UVIS observations from the
HDUV program (Oesch et al.\ 2018b) as well as the previously existing
WFC3/UVIS observations from the WFC3/IR Early Release Science (ERS)
and UVUDF programs (Windhorst et al.\ 2011; Teplitz et al.\ 2013)
allow us to extend the Bouwens et al.\ (2015) study of the $UV$ LF
down to $z\sim2$, while adding valuable statistics and leverage at the
bright and faint ends.

In addition, through inclusion of observations from the Hubble
Frontier Fields program (Lotz et al.\ 2017), we can further refine our
earlier determinations of the $UV$ LF at $z\sim4$-10 published in
Bouwens et al.\ (2015).  Importantly, the HFF parallel data probe
$\sim$1 mag fainter than the CANDELS data set, providing us with
probes of the volume density of galaxies at magnitude levels
intermediate between the CANDELS and XDF/HUDF regimes.

In the present determinations of the $UV$ LF, we expressly focus on
blank field search results for $z\sim2$-9 galaxies.  We exclude search
results behind lensing clusters to ensure that the present $UV$ LF
determinations are only impacted by systematic errors specific to
blank field studies (Bouwens et al.\ 2017a, Bouwens et al.\ 2017b;
Atek et al.\ 2018).  In a follow-up paper (Bouwens et al.\ 2021, in
prep), we will provide separate determinations of the $UV$ LF using
observations over the Hubble Frontier Fields clusters, and then we
will compare the LF results from the lensing fields with the blank
fields and test for consistency.

We now present a plan for this paper.  \S2 provides a brief
description of the data sets used in this study, our procedure for
deriving the photometry, and the selection criteria utilized in this
study.  In \S3, we summarize our procedure for deriving LF results,
while also presenting our new UV LF results.  In \S4, we discuss the
new trends we find and compare our new LF results with previous
results in the literature.  Finally, \S5 summarizes our results.

For convenience, we quote results in terms of the approximate
characteristic luminosity $L_{z=3}^{*}$ derived at $z\sim3$ by Steidel
et al.\ (1999), Reddy \& Steidel (2009), and many other studies.  We
refer to the {\it HST} F225W, F275W, F336W, F435W, F606W, F600LP,
F775W, F814W, F850LP, F098M, F105W, F125W, F140W, and F160W bands as
$UV_{225}$, $UV_{275}$, $U_{336}$, $B_{435}$, $V_{606}$, $V_{600}$,
$i_{775}$, $I_{814}$, $z_{850}$, $Y_{098}$, $Y_{105}$, $J_{125}$,
$JH_{140}$, and $H_{160}$, respectively, for simplicity.  The standard
concordance cosmology $\Omega_0 = 0.3$, $\Omega_{\Lambda} = 0.7$, and
$H_0 = 70\,\textrm{km/s/Mpc}$ is assumed for consistency with previous
LF studies.  All magnitudes are in the AB system (Oke \& Gunn 1983).

\begin{figure*}
\epsscale{1.14} \plotone{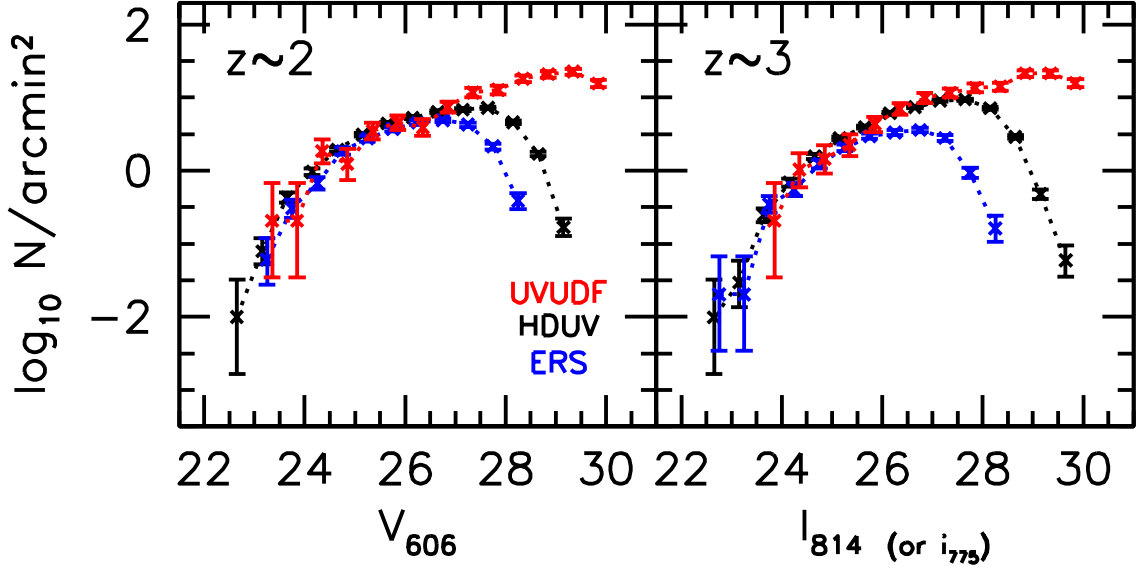}
\caption{Surface densities of the candidate $z\sim2$ and $z\sim3$
  galaxies for the three search fields considered in this analysis,
  i.e., ERS (\textit{blue points}), HDUV (\textit{black points}), and
  the UVUDF (\textit{red points}).  Surface densities are presented as
  a function of the $V_{606}$ and $I_{814}$ band magnitudes that
  provide the best measure of the rest-frame $UV$ flux of galaxies at
  1600$\AA$ for our $z\sim2$ and $z\sim3$ selections, respectively.
  For the UVUDF $z\sim3$ results, the $i_{775}$ band magnitudes are
  presented here instead (due to the significantly greater depth of
  the $i_{775}$-band data).  A slight horizontal offset of the points
  relative to each other has been applied for clarity.  The onset of
  incompleteness in our different samples is clearly seen in the
  observed decrease in surface density of sources near the magnitude
  limit.  We do not make use of the faintest sources in each search
  field, i.e., $V_{606}/i_{775}/I_{814}$ magnitudes fainter than 26.5,
  28.0, and 29.0 for the ERS, HDUV, and UVUDF fields, respectively,
  given the large uncertainties in the completeness (and
  contamination) corrections.\label{fig:surfdens}}
\end{figure*}

\section{Data Sets and Catalogues}

\subsection{HDUV + ERS}

The primary data for our $z=2$-3 LF results are the sensitive near-UV
observations obtained over a $\sim$94 arcmin$^2$ area within the
GOODS-South and GOODS-North fields using the HDUV program (Oesch et
al.\ 2018b).  For a description of the characteristics and reduction
of those data, we refer the interested reader to Oesch et
al.\ (2018b).  Optical and near-IR observations over this field were
obtained by making use of the v1.0 Hubble Legacy Field (HLF:
Illingworth et al.\ 2016; Whitaker et al.\ 2019; G.D. Illingworth et
al.\ 2021, in prep) reductions.  The HLF reductions constitute a
comprehensive reduction of all the archival optical/ACS +
near-IR/WFC3/IR observations over the GOODS-South and GOODS-North
fields.

For our $z=2$-3 selections and LF results, we also make use of the
WFC3/UVIS $UV_{225} UV_{275} U_{336}$ observations that were part of
the WFC3 ERS program over the GOODS South field.  These data cover
$\sim$50 arcmin$^2$.  The ERS observations, together with the HDUV
observations, cover an area of 143 arcmin$^2$ in total.  First
selections of $z\sim2$-3 galaxies from those data sets and rest-frame
$UV$ LF results were obtained by Hathi et al.\ (2010) and Oesch et
al.\ (2010).  As in the case of the HDUV data, we make use of the
reduction of optical and near-IR observations over the ERS area from
the HLF program.

\begin{deluxetable*}{lrrrrrrrrrr}
\tablewidth{0pt}
\tablecolumns{11}
\tabletypesize{\footnotesize}
\tablecaption{Total number of sources in the $z\sim2$, $z\sim3$, $z\sim4$, $z\sim5$, $z\sim6$, $z\sim7$, $z\sim8$, $z\sim9$, and $z\sim10$ samples from this paper and Oesch et al.\ 2018a\label{tab:sampnumbers}}
\tablehead{
\colhead{} & \colhead{Area} &
\colhead{$z\sim2$} & \colhead{$z\sim3$} & \colhead{$z\sim4$} & \colhead{$z\sim5$} & \colhead{$z\sim6$} & \colhead{$z\sim7$} & \colhead{$z\sim8$} & \colhead{$z\sim9$} & \colhead{$z\sim10$}\\
\colhead{Field} & \colhead{(arcmin$^2$)} & \colhead{\#} & \colhead{\#} & \colhead{\#} & \colhead{\#} & \colhead{\#} & \colhead{\#} & \colhead{\#} & \colhead{\#}}
\startdata
\multicolumn{11}{c}{From Bouwens et al.\ 2015 and 2016 (see also Oesch et al. 2013, 2014)}\\
HUDF/XDF & 4.7  & --- & --- & 357 & 153 & 97 & 57 & 29 & --- & ---\\
HUDF09-1 & 4.7  & --- & --- & --- & 91 & 38 & 22 & 18 & --- & 0\\
HUDF09-2 & 4.7  & --- & --- & 147 & 77 & 32 & 23 & 15 & --- & 0\\
CANDELS-GS-DEEP & 64.5  & --- & --- & 1590 & 471 & 198 & 77 & 26 & 2 & 1\\
CANDELS-GS-WIDE & 34.2  & --- & --- & 451 & 117 & 43 & 5 & 3 & 0 & 0\\
ERS & 40.5  & --- & --- & 815 & 205 & 61 & 46 & 5 & 2 & 0\\
CANDELS-GN-DEEP & 68.3  & --- & --- & 1628 & 634 & 188 & 134 & 51 & 1 & 2\\
CANDELS-GN-WIDE & 65.4  & --- & --- & 871 & 282 & 69 & 39 & 18 & 0 & 1\\
CANDELS-UDS & 151.2  & --- & --- & --- & 270 & 33 & 18 & 6 & 1 & 0\\
CANDELS-COSMOS & 151.9  & --- & --- & --- & 320 & 48 & 15 & 9 & 1 & 0\\
CANDELS-EGS & 150.7  & --- & --- & --- & 381 & 50 & 43 & 9 & 2 & 1\\
BORG/HIPPIES & 218.3  & --- & --- & --- & --- & --- & --- & 23 & --- & ---\\
\\
\multicolumn{11}{c}{HDUV + ERS + UVUDF (This Work [\S2.6])}\\
HDUV-GOODS-S (+ UVUDF) & 43.5  & 2127 & 2454 & --- & --- & --- & --- & --- & --- & ---\\
ERS & 49.2  & 1252 & 1055 & --- & --- & --- & --- & --- & --- & ---\\
HDUV-GOODS-N & 57.6  & 2387 & 2823 & --- & --- & --- & --- & --- & --- & ---\\
\\
\multicolumn{11}{c}{From CANDELS COSMOS/UDS/EGS Fields (Bouwens et al. 2019 and This Work (\S2.8])}\\
CANDELS-UDS & 45.3  & --- & --- & --- & --- & --- & --- & --- & 1 & 0\\
CANDELS-COSMOS & 48.7  & --- & --- & --- & --- & --- & --- & --- & 0 & 0\\
CANDELS-EGS & 53.4  & --- & --- & --- & --- & --- & --- & --- & 4 & 0\\
\\
\multicolumn{11}{c}{Hubble Frontier Fields Parallels (This Work [\S2.7,2.8] + Oesch et al. 2018a)}\\
Abell 2744-Par & 4.9  & --- & --- & 226 & 67 & 20 & 11 & 4 & 3 & 0\\
MACS0416-Par & 4.9  & --- & --- & 266 & 71 & 25 & 19 & 4 & 3 & 0\\
MACS0717-Par & 4.9  & --- & --- & 214 & 55 & 41 & 21 & 10 & 0 & 0\\
MACS1149-Par & 4.9  & --- & --- & 234 & 76 & 36 & 31 & 6 & 1 & 0\\
Abell S1063-Par & 4.9  & --- & --- & 231 & 79 & 40 & 20 & 7 & 2 & 0\\
Abell 370-Par & 4.9  & --- & --- & 210 & 100 & 47 & 20 & 3 & 4 & 2\\
HFF Total & 29.4 & --- & --- & 1381 & 448 & 209 & 122 & 34 & 13 & 2\\
\\
\multicolumn{11}{c}{Hubble Ultra Deep Field + Parallels (This Work [\S2.8] + Oesch et al. 2018a)}\\
HUDF/XDF & 4.9  & --- & --- & --- & --- & --- & --- & --- & 4 & 1\\
HUDF09-1 & 4.9  & --- & --- & --- & --- & --- & --- & --- & 0 & 0\\
HUDF09-2 & 4.9  & --- & --- & --- & --- & --- & --- & --- & 2 & 0\\
Total & 1135.9 & 5766 & 6332 & 7240 & 3449 & 1066 & 601 & 246 & 33 & 8
\enddata
\end{deluxetable*}

Figure~\ref{fig:layout} shows the layout of the WFC3/UVIS observations
from the HDUV and ERS fields over the GOODS-South and GOODS-North
fields.

\subsection{UVUDF/XDF}

We also made use of near-UV, optical, and near-IR observations over
the HUDF from the UVUDF program (Teplitz et al.\ 2013), optical ACS
HUDF program (Beckwith et al.\ 2016), HUDF09/HUDF12 programs (Bouwens
et al.\ 2011; Ellis et al.\ 2013), and any other {\it HST}
observations that have been taken over the HUDF/XDF.  Illingworth et
al.\ (2013) combined all existing optical and near-IR observations
over the HUDF (including many archival observations) into an
especially deep reduction called the eXtreme Deep field (XDF).  The
XDF optical reductions include all ACS and WFC3/IR data on the HUDF
through 2013 and are $\sim$0.1-0.2 mag deeper than the Beckwith et
al.\ (2006) reductions of the optical ACS data.

We make use of the v2.0 reductions of the epoch 3 WFC3/UVIS data over
the HUDF acquired in post-flash mode (to cope with CTE degradation:
see Rafelski et
al.\ 2015).\footnote{https://archive.stsci.edu/prepds/uvudf/}
Observations for epoch 3 of the UVUDF program were divided equally
across the F225W, F275W, and F336W bands, with 15 orbits of time
allocated to each band.  The $5\sigma$ depths we measure for the
epoch-3 UVUDF data in $0.4''$-diameter apertures are 27.1, 27.2, and
27.8 mag, respectively.  No use was made of the first 45 orbits of
data from the UVUDF program, given the impact of CTE degration on
those data which were acquired without post-flash (see Teplitz et
al.\ 2013).

\subsection{Parallel Fields to the Hubble Ultra Deep Field}

Another valuable data set we use for our $z\sim9$ search are the two
flanking fields to the HUDF, i.e., HUDF09-1 and HUDF09-2 (Oesch et
al.\ 2007; Bouwens et al.\ 2011) where sensitive observations have
been obtained with both ACS and WFC3/IR.  These observations could be
obtained efficiently due to simultaneous observing programs over the
HUDF and due to the parallel observing capabilities of {\it HST}.  A
total of 8, 12, and 13 orbits in the $Y_{105}$, $J_{125}$, and
$H_{160}$ bands, respectively, were obtained over HUDF09-1 parallel
field, while 11, 18, and 19 orbits in the $Y_{105}$, $J_{125}$,
$H_{160}$ bands, respectively, were obtained over HUDF09-2 parallel
field.  Very deep ($>$100 orbits) optical data in the
$V_{606}i_{775}I_{814}z_{850}$ bands also exist over these two fields
from the HUDF05, HUDF09, HUDF12, and other programs (Oesch et
al.\ 2007; Bouwens et al.\ 2011; Ellis et al.\ 2013).

\subsection{Hubble Frontier Fields Parallels}

In addition to the data already utilized in Bouwens et al.\ (2015) and
Bouwens et al.\ (2016) for blank-field LF results at $z=4$, 5, 6, 7,
8, and 9, we also add the sensitive optical and near-IR observations
obtained over six deep parallel fields from the HFF program (Coe et
al.\ 2015; Lotz et al.\ 2017).  These deep parallel fields supplement
the deep optical and near-IR observations obtained by the HFF program
over the centers of six different clusters (Abell 2744, MACS0416,
MACS0717, MACS1149, Abell 370, and Abell S1063) and are separated from
the cluster centers by $\sim$8 arcmin.  70 orbits of optical ACS
observations (18, 10, and 42 in the F435W, F606W, and F814W bands,
respectively) and 70 orbits of WFC3/IR observations (24, 12, 10, and
24 in the F105W, F125W, F140W, and F160W bands, respectively) were
invested in observations of each parallel field.  We made use of the
v1.0 reductions of these observations made publicly available by the
HFF team (Koekemoer et al.\ 2014).

In addition to making use of the available {\it HST} observations, we
also made use of the $\sim$50-80 hours of {\it Spitzer}/IRAC
observations over the parallel fields to the HFF clusters to allow for
the selection of galaxies to $z\sim9$.  The available observations
were drizzled together to construct sensitive mosaics of each cluster
at $\sim$3-5 microns (as performed by Labb{\'e} et al.\ 2015 and
Stefanon et al.\ 2020).

\begin{deluxetable*}{ccccccc}
\tablewidth{0pt} \tablecolumns{11} \tabletypesize{\footnotesize}
\tablecaption{A complete list of the sources included in the $z\sim2$,
  $z\sim3$, $z\sim 4$, $z\sim 5$, $z\sim 6$, $z\sim 7$, $z\sim 8$,
  $z\sim9$, and $z\sim 10$ samples from the present selection and that
  of Oesch et al.\ 2018a\tablenotemark{*}\label{tab:catalog}}
\tablehead{ \colhead{ID} & \colhead{R.A.} & \colhead{Dec} &
  \colhead{$m_{AB}$\tablenotemark{a}} & \colhead{Sample\tablenotemark{b}} &
  \colhead{Data Set\tablenotemark{c}} &
  \colhead{$z_{phot}$\tablenotemark{d,e}}} \startdata XDFB-2384848214
& 03:32:38.49 & $-$27:48:21.4 & 27.77 & 4 & 1 & 3.49
\\ XDFB-2384248186 & 03:32:38.42 & $-$27:48:18.7 & 29.18 & 4 & 1 &
3.82 \\ XDFB-2376648168 & 03:32:37.66 & $-$27:48:16.9 & 28.61 & 4 & 1
& 4.01 \\ XDFB-2385948162 & 03:32:38.60 & $-$27:48:16.2 & 28.04 & 4 &
1 & 4.16 \\ XDFB-2382548139 & 03:32:38.26 & $-$27:48:13.9 & 28.18 & 4
& 1 & 4.37 \\ XDFB-2394448134 & 03:32:39.45 & $-$27:48:13.4 & 26.40 &
4 & 1 & 3.58 \\ XDFB-2381448127 & 03:32:38.14 & $-$27:48:12.7 & 28.58
& 4 & 1 & 3.68 \\ XDFB-2390248129 & 03:32:39.03 & $-$27:48:13.0 &
27.99 & 4 & 1 & 3.91 \\ XDFB-2379348121 & 03:32:37.93 & $-$27:48:12.1
& 27.45 & 4 & 1 & 4.11 \\ XDFB-2378848108 & 03:32:37.88 &
$-$27:48:10.9 & 30.13 & 4 & 1 & 3.72 \enddata
\tablenotetext{*}{Table~\ref{tab:catalog} is published in its entirety
  in the electronic edition of the Astrophysical Journal.  A portion
  is shown here for guidance regarding its form and content.}
\tablenotetext{a}{Apparent magnitude in $V_{606}$, $I_{814}$, and $H_{160}$ 
band for galaxies in the $z\sim2$, $z\sim3$, and $z\sim4$-10 samples, respectively.  
Apparent magnitudes are in the $i_{775}$ band for $z\sim3$ sources over the UVUDF.}
\tablenotetext{b}{The mean redshift of the sample in which the source
  was included for the purposes of deriving LFs.}
\tablenotetext{c}{The data set from which the source was selected: 1 =
  HUDF/XDF, 2 = HUDF09-1, 3 = HUDF09-2, 4 = ERS, 5 = CANDELS-GS, 6 =
  CANDELS-GN, 7 = CANDELS-UDS, 8 = CANDELS-COSMOS, 9 = CANDELS-EGS, 10
  = BoRG/HIPPIES or other pure-parallel programs, 11 = Abell2744-Par,
  12 = MACS0416-Par, 13 = MACS0717-Par, 14 = MACS1149-Par, 15 = Abell
  S1063, and 16 = Abell 370} \tablenotetext{d}{Most likely redshift in
  the range $z=2.5$-11 as derived using the EAZY photometric redshift
  code (Brammer et al.\ 2008) using the same templates as discussed in
  \S3.2.3.}  \tablenotetext{e}{``*'' indicates that for a flat
  redshift prior, the EAZY photometric redshift code (Brammer et
  al.\ 2008) estimates that this source shows at least a 68\%
  probability for having a redshift significantly lower than the
  nominal low-redshift limit for a sample, i.e., $z<0.8$, $z<1.5$,
  $z<2.5$, $z<3.5$, $z<4.4$, $z<5.4$, $z<6.3$, and $z<8$ for candidate
  $z\sim2$, $z\sim3$, $z\sim4$, $z\sim5$, $z\sim6$, $z\sim7$,
  $z\sim8$, and $z\sim10$ galaxies, respectively.}
\end{deluxetable*}

\subsection{Source Detection and Photometry}

Our procedures for pursuing source detection and photometry are very
similar to most of our previous work (e.g., Bouwens et al.\ 2011,
2015).  We use the SExtractor software (Bertin \& Arnouts 1996) to
handle source detection and photometry.  We run the SExtractor
software in dual-image mode, with the detection image taken to equal
the square root of $\chi^2$ image (Szalay et al.\ 1999: similar to a
coadded image) constructed from the $V_{606} i_{775} I_{814} z_{850}$
images for our $z\sim2$-3 selections, constructed from the $Y_{105}
J_{125} JH_{140} H_{160}$ images for our $z\sim4$-7 selections,
$J_{125} JH_{140} H_{160}$ images for our $z\sim8$ selections, and
$JH_{140}$ and $H_{160}$ images for our $z\sim9$ selections.  Color
measurements are made in small scalable apertures (Kron [1980] factor
of 1.2), after PSF-matching the observations to the $z_{850}$ band (if
the color measurement only includes the optical bands) or the
$H_{160}$ band (if the color measurement includes a near-infrared
band).

Measurements of the total magnitude are made by correcting the
smaller-scalable aperture flux measurements to account for the excess
flux measured in the larger-scalable apertures relative to the
smaller-scalable apertures and also for the light on the wings on the
PSF (typically a $\sim$0.15-0.25 mag correction) using the tabulated
values of the encircled energy distributions (Dressel et al. 2012).

For $z\sim9$ selections, only the {\it HST} $JH_{140}$ and $H_{160}$
probe the spectral slope of galaxies redward of the Lyman-break
providing us with very limited leverage to distinguish bona-fide
star-forming galaxies at $z\sim9$ from lower-redshift interlopers.
Therefore, for our $z$$\sim$9 selections, we also derive fluxes for
individual sources at $\sim$3.6$\mu$m and 4.5$\mu$m using the
\textsc{mophongo} software (Labb{\'e} et al.\ 2006, 2010a, 2010b,
2013, 2015).  Deriving fluxes for sources in the $3.6\mu$m and
$4.5\mu$m bands is challenging due to the broad PSF of the {\it
  Spitzer}/IRAC data, which causes light from neighboring sources to
blend together on the images.  To overcome these issues,
\textsc{mophongo} uses the high spatial resolution {\it HST} data to
create template images of each source in the lower spatial resolution
{\it Spitzer}/IRAC data and then the fluxes of the source and its
neighbors is varied to obtain the best fit.  The model profiles of the
neighboring sources is then subtracted from the image, and then the
flux of the source is measured in $1.8''$-diameter apertures.  These
fluxes are then extrapolated to total based on the model profile of
the source convolved with the PSF.

In selecting candidate $z=2$-9 galaxies, we required the candidate
galaxies in our $z\sim2$-3, $z\sim4$-7, $z\sim8$, and $z\sim9$ samples
to show a S/N of 5.5, 5.5, 6, and 6.5, respectively, in the $\chi^2$
images used to detect sources.  Sources which correspond to
diffraction spikes, are the clear result of an elevated background
around a bright source (e.g., for a bright elliptical galaxy), or
correspond to other artifacts in the data are removed by visual
inspection.

We clean the sample by removing all bright ($H_{160,AB}<27$) sources
with SExtractor stellarity parameters in excess of 0.9, i.e.,
star-like.  SExtractor stellarity parameters of 0 and 1 correspond to
extended and point sources, respectively.  We also removed all sources
with whose SExtractor stellarity parameter is in excess of 0.6 and
whose {\it HST} photometry is much better fit with an SED of a
low-mass star ($\Delta \chi^2 > 2$) from the SpeX library (Burgasser
et al.\ 2004) than with a linear combination of galaxy templates from
EAZY (Brammer et al.\ 2008).

\subsection{Selection of $z=2$-3 Galaxies\label{sec:z23}}

As in our own previous searches for $z\sim2$-9 galaxies (e.g., Oesch
et al.\ 2010; Oesch et al.\ 2013; Bouwens et al.\ 2015), we required
sources to satisfy Lyman-break-like criteria for inclusion in our
samples.  In fact, spectroscopic follow-up work has demonstrated that
Lyman-break-like color-color criteria provide a very efficient way of
identifying $z\sim3$-8 Lyman-break galaxies (e.g., Steidel et
al.\ 1999; Steidel et al.\ 2003; Vanzella et al.\ 2009; Stark et
al.\ 2010; Ono et al.\ 2012; Finkelstein et al.\ 2013; Oesch et
al.\ 2015; Zitrin et al.\ 2015; Hashimoto et al.\ 2018).

In our selection of galaxies for our $z\sim2$ and $z\sim3$ samples
from the HDUV and ERS data, we first apply the following criteria to
our source catalogs:
\begin{eqnarray*}
(UV_{275}-B_{435}>1)\wedge \\
((V_{606}-z_{850}<0.5)\vee \\
((UV_{275}-B_{435}>2(V_{606}-z_{850}))\wedge\\
(V_{606}-z_{850}<1.0)))
\end{eqnarray*}
or
\begin{eqnarray*}
(UV_{336}-V_{606}>1)\wedge \\
((V_{606}-z_{850}<0.5)\vee \\
((UV_{336}-V_{606}>2(V_{606}-z_{850}))\wedge\\
(V_{606}-z_{850}<1.0)))\wedge
(\textrm{SN}(UV_{275})<2)
\end{eqnarray*}
where $\wedge$, $\vee$, and SN represents the logical \textbf{AND}
operation, the logical \textbf{OR} operation, and signal to noise
computed in small scalable apertures, respectively.  The fluxes of
sources not detected are set to the $1\sigma$ upper limits on the flux
in the undetected band.

\begin{figure*}
\epsscale{1.17}
\plotone{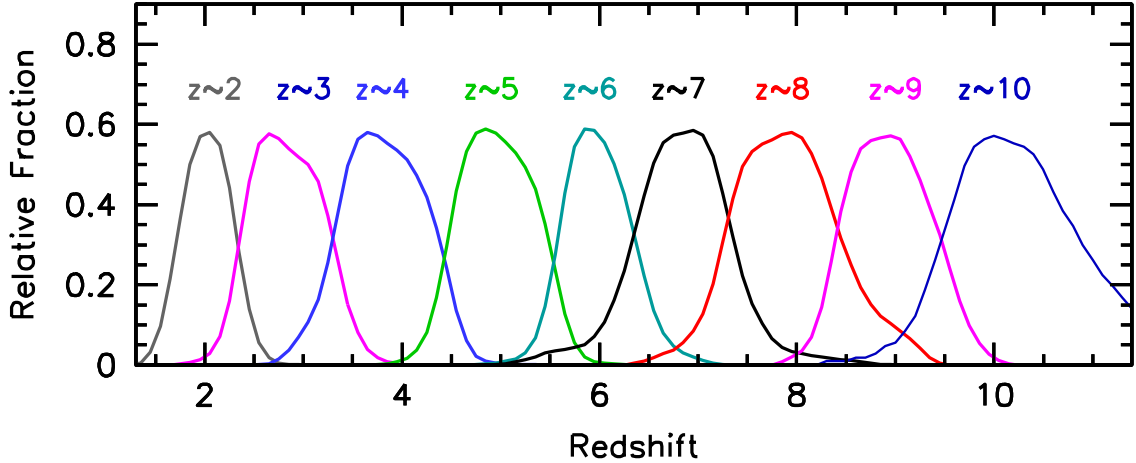}
\caption{Shown is the approximate redshift distribution expected for
  sources in our selections of $z\sim2$, $z\sim3$, $z\sim4$, $z\sim5$,
  $z\sim6$, $z\sim7$, $z\sim8$, and $z\sim9$ galaxies with the grey,
  magenta, blue, green, cyan, black, red, and magenta lines,
  respectively.  The expected redshift distributions shown here are
  based on our HDUV and HUDF/XDF selection volume simulation results
  at $z\sim2$-3 and $z\sim4$-9, respectively.  The dark blue line
  shows the expected redshift distribution for the $z\sim10$ selection
  from the companion study of Oesch et al.\ (2018a).  The precise
  redshift distribution exhibits a modest dependence on the available
  {\it HST} passbands for a data set, as illustrated e.g. in Figure 4
  of Bouwens et al.\ (2015).\label{fig:zdist}}
\end{figure*}

We then make use of the photometric redshift software EAZY (Brammer et
al.\ 2008) to determine the redshift likelihood distribution for each
source.  Consideration was made of the photometry we derived in the
WFC3/UVIS (UV$_{275}$, $U_{336}$), ACS ($B_{435}$, $V_{606}$,
$i_{775}$, $I_{814}$, $z_{850}$), and WFC3/IR ($Y_{098}$, $Y_{105}$,
$J_{125}$, $JH_{140}$, and $H_{160}$) bands.  The SED templates we used
were the EAZY\_v1.0 set supplemented by SED templates from the Galaxy
Evolutionary Synthesis Models (GALEV: Kotulla et al.\ 2009).  Nebular
continuum and emission lines were added to the later templates using
the Anders \& Fritze-v. Alvensleben (2003) prescription, a $0.2
Z_{\odot}$ metallicity, and a rest-frame EW for H$\alpha$ of 1300\AA.
To allow for possible systematics in our photometry and differences
between the observed and model SEDs, we assume an additional 7\%
uncertainty in our flux measurements when deriving photometric
redshifts with EAZY.

\begin{figure*}
\epsscale{0.8}
\plotone{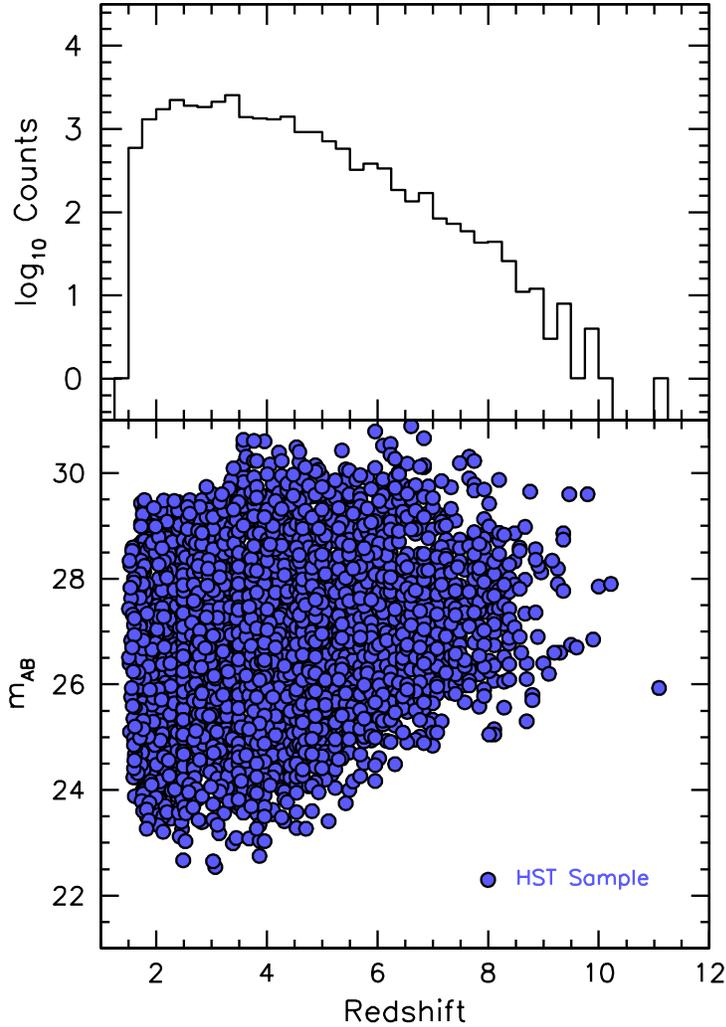}
\caption{(\textit{upper}) Histogram of the \# of sources vs. redshift
  for the {\it HST} selections considered here.  (\textit{lower})
  Redshift vs. apparent magnitudes (\textit{blue filled circles}) for
  all sources in the present {\it HST} samples (and those of Oesch et
  al.\ 2018a).  The source at $z\sim11.1$ and with an $H_{160,AB}$
  magnitude of 25.9 mag is GN-z11 (Bouwens et al.\ 2010; Oesch et
  al.\ 2014, 2016).\label{fig:twod}}
\end{figure*}

For selection, we additionally required that $>$65\% of the integrated
probability in the photometric redshift likelihood distribution lie at
$>$1.2 and for the best-fit $\chi^2$ be less than 25 (equivalent to
$\chi_{reduced} ^2 \lesssim 2.5$) to include sources where we can
obtain a reasonable SED fit to the photometry.  Sources where the
best-fit photometric redshift lie in the range $z=1.5$-2.5 and
$z=2.5$-3.5 are placed in our $z\sim2$ and $z\sim3$ samples,
respectively.

\subsection{Selection of $z=4$-8 Galaxies}

As in previous work (Bouwens et al.\ 2015), we select $z=4$-9 galaxies
from the HFF parallel fields using Lyman-break color criteria.
Sources in our $z\sim4$ samples are selected following these criteria:
\begin{eqnarray*}
(B_{435}-V_{606}>1.0)\wedge(I_{814}-J_{125}<1.0)\wedge\\
(B_{435}-V_{606}>1.0+1.77(I_{814}-J_{125})\wedge\\
\textrm{[not in}\,z\sim5\,\,\textrm{samples]}
\end{eqnarray*}

Our $z\sim5$ samples are selected using the following color criteria:
\begin{eqnarray*}
(V_{606}-I_{814}>1.2)\wedge\\
(V_{606}-I_{814}>1.32+1.2(Y_{105}-H_{160}))\wedge\\
(Y_{105}-H_{160}<0.9)\wedge(\textrm{SN}(B)<2)\wedge \\
\textrm{[not in}\,z\sim6-7\,\,\textrm{samples]}
\end{eqnarray*}

For our $z\sim6$ and $z\sim7$ samples, we select sources using the
following color criteria:
\begin{eqnarray*}
(I_{814}-Y_{105}>0.6)\wedge(Y_{105}-H_{160}<0.45)\wedge\\
(I_{814}-Y_{105}>0.6(Y_{105}-H_{160}))\wedge\\
(Y_{105}-H_{160}<0.52+0.75(J_{125}-H_{160})\wedge\\
\textrm{SN}(B_{435}<2)\wedge\\
((\chi_{opt} ^2(B_{435},V_{606})<2)\vee(V_{606}-Y_{105}>2.5))\wedge\\
\textrm{[not in}\,z\sim 8\,\,\textrm{samples]}
\end{eqnarray*}
where $\chi_{opt} ^2$ is calculated as follows $\chi_{opt} ^2 =
\Sigma_{i} \textrm{SGN}(f_{i}) (f_{i}/\sigma_{i})^2$ where $f_{i}$ is
the flux in band $i$ in a consistent aperture, $\sigma_i$ is the
uncertainty in this flux, and SGN($f_{i}$) is equal to 1 if $f_{i}>0$
and $-1$ if $f_{i}<0$ (see Bouwens et al.\ 2011).

For our $z\sim8$ selection, we apply the following criteria:
\begin{eqnarray*}
(Y_{105}-J_{125}>0.45)\wedge\\
(Y_{105}-J_{125}>0.525+0.75(J_{125}-H_{160})\wedge\\
(J_{125}-H_{160}<0.5)\wedge (\chi_{opt,0.35''} ^2 < 4)\wedge \\
(\chi_{opt,Kron} ^2 < 4)\wedge(\chi_{opt,0.2''} ^2 < 4)\wedge\\
\textrm{[not in}\,z\sim 9\,\,\textrm{samples]}
\end{eqnarray*}
Sources in our $z\sim8$ sample must have a $\chi^2$ statistic less
than 4 (i.e., $<$2$\sigma$ detection) combining the $B_{435}$,
$V_{606}$, and $I_{814}$-band flux measurements in both small scalable
apertures and fixed 0.35$''$-diameter apertures.

We divide the $z\sim6$-7 selection into $z\sim6$ and $z\sim7$ samples
using the photometric redshift we compute for individual sources using
the EAZY photometric redshift software (Brammer et al.\ 2008). Sources
with a photometric redshift $z<6.3$ are assigned to our $z\sim6$
sample provided that the fractional likelihood of the source lying at
$z<4.3$ is $<$35\%, whereas sources with a photometric redshift
$z>6.3$ are assigned to our $z\sim7$ selection.  Sources in our
$z\sim7$ sample must have a $\chi^2$ statistic less than 4 (i.e.,
$<$2$\sigma$ detection) combining the $B_{435}$ and $V_{606}$ flux
measurements in small scalable apertures and fixed 0.35$''$-diameter
apertures.

\subsection{Selection of $z\sim9$ Galaxies}

In selecting candidate $z\sim9$ galaxies from the HFF parallel fields
and the XDF, both of which have deep $JH_{140}$ observations, we make
use fo the following color criteria to identify candidate $z\gtrsim9$
galaxies:
\begin{eqnarray*}
((Y_{105}-H_{160})+2(J_{125}-JH_{140})>1.5)\wedge\\
((Y_{105}-H_{160})+2(J_{125}-JH_{140})>\\
~~~~~~~1.5+1.4(JH_{140}-H_{160}))\wedge\\
((Y_{105}-H_{160})+(Y_{105}-JH_{140})>2)\wedge\\
(JH_{140}-H_{160}<0.5)\wedge
(\chi_{opt,0.35''} ^2 < 4)\wedge \\
(\chi_{opt,Kron} ^2 < 4)\wedge(\chi_{opt,0.2''} ^2 < 4)
\end{eqnarray*}
where $\chi_{opt,0.35''} ^2$, $\chi_{opt,Kron} ^2$, and
$\chi_{opt,0.2''} ^2$, respectively, represent the ``$\chi^2$''
statistic computed from the optical fluxes in 0.35$''$-diameter
apertures, small-scalable Kron apertures, and small 0.2$''$-diameter
apertures (before PSF-matching the optical data to the lower
resolution near-IR data).

For the two deep parallel fields to the HUDF, HUDF09-1 and HUDF09-2,
deep $JH_{140}$-band data are not available, and so we utilize the
following color criteria:
\begin{eqnarray*}
((Y_{105}-H_{160})+2(J_{125}-H_{160})>1.5)\wedge\\
(J_{125}-H_{160} < 1.2)\wedge (\chi_{opt,0.35''} ^2 < 4)\wedge \\
(\chi_{opt,Kron} ^2 < 4)\wedge(\chi_{opt,0.2''} ^2 < 4)
\end{eqnarray*}
In cases of a non-detection, the measured fluxes are set to their
$1\sigma$ upper limits for the purposes of deriving measured colors to
apply the above criteria.

Our $z\sim9$ selection criteria are modified from those presented in
Oesch et al.\ (2013).  This is in an attempt to contrast the
``average'' flux information in the $Y_{105}$ and $J_{125}$ bands and
the ``average'' flux information in the $JH_{140}$ and $H_{160}$ bands
to measure the size of the apparent break in the spectrum of candidate
$z\sim9$ galaxies.  In computing the $\chi^2$ statistic for sources in
our $z\sim9$ selections, we included the fluxes in all optical ACS
bands blueward of $Y_{105}$.

To maximize the robustness of the sources in our selection, we also
made use of the {\it Spitzer}/IRAC observations of the $z\sim9$
candidates to examine the color of the sources redward of the nominal
Lyman break.  We considered both our own photometry on each candidate
and that from Shipley et al.\ (2018) for those sources falling within
the HFF parallel fields.  Given the challenges of obtaining {\it
  Spitzer}/IRAC flux measurements in the presence of source crowding,
we only excluded sources if they showed at least a $3\sigma$ detection
both from our own photometry and that from Shipley et al.\ (2018) and
if the source showed a $H_{160}-[3.6]$ color redder than 0.7 mag.

Finally, sources are required to have a best-fit photometric redshift
calculated with EAZY between $z=8.4$ and $z=9.5$ and to have $>$70\%
of the redshift likelihood distribution above $z\sim7$.  We used the
same SED template set to compute this redshift likelihood distribution
as we used in \S\ref{sec:z23}.

Our $z\sim9$ selection also includes sources identified over the five
CANDELS fields and ERS field, a 874 arcmin$^2$ area.  While we have
already provided an extensive description of this selection in Bouwens
et al.\ (2019), some additional $Y_{098}$ and $JH_{140}$ imaging has
become available on $z\sim9$ candidates from that selection thanks to
observations from HST programs 15103 (PI: de Barros) and 15862 (PI:
Finkelstein).  $Y_{098}$ and $Y_{105}$-band observations from those
program further confirm the nature of COS910-1, EGS910-9, and
EGS910-10, with estimated $P(z>8)$ probabilities of 0.97, 0.75, and
1.0, respectively, and strengthen the case that EGS910-15 is at $z>8$,
with $P(z>8)$ being 0.56.

Our previous $z\sim4$-10 LF study (Bouwens et al.\ 2015) made no use
of a separate $z\sim9$ selection, and therefore many $z\sim9$ galaxies
might have been included in their $z\sim7$ and $z\sim8$ samples (which
included a tail extending up to $z\sim9$).  We therefore inspected the
$z\sim7$ and $z\sim8$ samples from Bouwens et al.\ (2015) to search
for overlap with our new $z\sim9$ samples and eliminated any sources
in common (10 candidates).  Additionally, we recomputed the selection
volumes from Bouwens et al.\ (2015) to explicitly exclude sources that
would also satisfy the present $z\sim9$ selection criteria.

\subsection{Derived Samples of $z\sim2$-9 Galaxies}

Applying our selection criteria to the WFC3/UVIS + optical ACS +
WFC3/IR observations over the GOODS South and GOODS North fields, we
identify a total of 5766 $z\sim2$ galaxies and 6332 $z\sim3$ galaxies.
The surface density of sources in our $z\sim2$ selections as a
function of the apparent magnitude in the $V_{606}$ band is shown in
Figure~\ref{fig:surfdens}, while the surface densities of our $z\sim3$
HDUV, UVUDF, and ERS samples are presented as a function of the
$I_{814}$, $i_{775}$, and $i_{775}$ band magnitudes, respectively.
These bands probe close to 1600\AA$\,$in the rest frame.

For comparison, Hathi et al.\ (2010) identified 66 $z\sim1.7$
$UV_{225}$ dropouts, 151 $z\sim2.1$ $UV_{275}$ dropouts, and 256
$z\sim2.7$ $U_{336}$ dropouts over the $\sim$50 arcmin$^2$ WFC3/IR ERS
field.  Meanwhile, Oesch et al.\ (2010) find 60 $UV_{225}$, 99
$UV_{275}$, and 403 $U_{336}$ dropouts over the same ERS field.
Combining the individual subsamples, Hathi et al.\ (2010) and Oesch et
al.\ (2010) find 473 $z\sim2$-3 and 562 $z\sim2$-3 galaxies over the
ERS field.  While we find a much larger number of sources over the
ERS, i.e., 2307 sources, the $z\sim2$ and $z\sim3$ selections of Hathi
et al.\ (2010) and Oesch et al.\ (2010) cut off approximately
$\sim$1.2 mag brightward of our selections due to their use of more
restrictive selection criteria.  If we similarly cut off our $z\sim2$
and $z\sim3$ selections at $\sim$25.5 mag and $\sim$26 mag, we find
876 $z\sim2$-3 galaxies, which is much more comparable to the numbers
in these previous selections.

Using the $\sim$7.3 arcmin$^2$ UVUDF data set, Mehta et al.\ (2017)
identify 852 $z\sim2$-3 galaxies.  This is fairly similar (just 20\%
smaller) than the 1069 $z\sim2$-3 galaxies we find over the same
field.  The surface density of $z\sim$2-3 galaxies in the Mehta et
al.\ (2017) samples, i.e., 120 galaxies arcmin$^{-2}$, is also
comparable, but 29\% larger, than the 93 galaxy arcmin$^{-2}$ surface
density we find over the HDUV fields.  It is because of the
combination of depth and area of the current UVUDF+UVUDF data sets,
i.e., $\sim$1-mag greater depth than ERS and 21 $\times$ larger area
than UVUDF+HDUV data sets relative to previousERS and UVUDF data sets
alone that the present $z\sim2$-3 samples are $>$10$\times$ larger
than the previous $z\sim2$-3 samples of Hathi et al.\ (2010), Oesch et
al.\ (2010), and Mehta et al.\ (2017).

For our $z\sim4$, $z\sim5$, $z\sim6$, $z\sim7$, $z\sim8$, and $z\sim9$
selections over the HFF parallel fields, a total of 1381, 448, 209,
122, 34, and 13 galaxies are identified.  Adding to these new sources
to those sources found in the Bouwens et al.\ (2015) samples from the
HUDF/XDF, the HUDF parallel fields, BoRG, and the five CANDELS fields,
our total samples of $z\sim4$, $z\sim5$, $z\sim6$, $z\sim7$, $z\sim8$,
$z\sim9$ are 7240, 3449, 1066, 601, 246, and 33.  These sources are in
addition to the 9 sources in the $z\sim 10$-11 samples of Oesch et
al.\ (2018a), for which this analysis was done in coordination.
Table~\ref{tab:sampnumbers} summarizes the number of sources in each
of the samples we consider.  The total size of our {\it HST} samples
at $z=2$-11 is 24741, 12643 of which are in the redshift range
$z\sim4$-11.  Table~\ref{tab:catalog} presents the complete catalog of
these sources, with coordinates, apparent magnitudes, and photometric
redshift estimates.

The expected redshift distributions for our $z\sim2$, $z\sim3$,
$z\sim4$, $z\sim5$, $z\sim6$, $z\sim7$, $z\sim8$, and $z\sim9$
selections are shown in Figure~\ref{fig:zdist}, along with the
redshift distribution for the $z\sim10$ selection from Oesch et
al.\ (2018).

The top panel of Figure~\ref{fig:twod} shows the total number of the
sources per unit $\Delta z\sim0.25$, while the lower panel shows the
full distribution of magnitudes and redshifts that sources in our
samples occupy.

\begin{deluxetable}{cc}
\tablewidth{0pt}
\tablecolumns{2}
\tabletypesize{\footnotesize}
\tablecaption{Magnification Factors Adopted for Each of the HFF Parallel Fields\label{tab:hffpar_mu}}
\tablehead{
\colhead{Field} & \colhead{Typical Magnification Factor $\mu$\tablenotemark{a}}}
\startdata
Abell 2744-Par &  1.16 \\
MACS0416-Par &  1.05 \\
MACS0717-Par &  1.16 \\
MACS1149-Par &  1.04 \\
Abell S1063-Par & 1.05 \\
Abell 370-Par &  1.10 
\enddata
\tablenotetext{a}{Estimated from the version 1 lensing models of Merten (2016).}
\end{deluxetable}

\section{Luminosity Function Results}

The purpose of the present section is to summarize our procedures for
deriving the $UV$ LFs at $z\sim2$, 3, 4, 5, 6, 7, 8, and 9.  The
present determinations leverage a variety of new data sets to improve
on the results obtained in Oesch et al.\ (2010), Bouwens et
al.\ (2015), and Bouwens et al.\ (2016).

Given that the present analysis aims to improve on earlier LF analyses
from Oesch et al.\ (2010), Bouwens et al.\ (2015), and Bouwens et
al.\ (2016), our new determinations still incorporate constraints from
earlier data sets, such as the HUDF, the two HUDF parallel fields, the
WFC3/IR ERS field, the five CANDELS fields, and 220 arcmin$^2$ in
search area from BoRG+HIPPIES utilized in Bouwens et al.\ (2015) and
Bouwens et al.\ (2016).  We also include $z\sim2$ and $z\sim3$ samples
from the UVUDF and WFC3/UVIS ERS fields.

Our procedure for deriving the selection volumes is identical to that
described in Appendix D of Bouwens et al.\ (2015) and involves
creating artificial sources with a variety of apparent magnitudes and
redshifts using our artificial redshifting code (Bouwens et al.\ 1998;
Bouwens et al.\ 2003), adding those sources to the observations, and
then selecting those sources in the same way as we do with the real
observations.  In creating artificial sources for our $z\sim2$-3
selection volume simulations, we used the pixel-by-pixel morphologies
of similar-luminosity galaxies from our $z\sim2$ HUDF samples, scaling
them in size as $(1+z)^{-1.2}$ and using the same $UV$ color
distribution as Bouwens et al.\ (2009) and Bouwens et al.\ (2014).
Selection volumes for our UVUDF selections are created in a similar
way, but computing photometric redshifts for the sources detected in
the simulations and applying our selection criteria to determine if a
simulated source is selected or not.  Since these simulations use
similar-luminosity $z\sim2$ galaxies from the HUDF to simulate
galaxies at $z\sim3$ or in shallower fields (like the HDUV or ERS
fields), they implicitly account for the size-luminosity relation.
Selection volumes for our $z\sim4$-9 samples follow the same
procedure, but start with a random ensemble of $z\sim4$ galaxies from
the HUDF/XDF data as selected by Bouwens et al.\ (2015).

\begin{figure}
\epsscale{1.15}
\plotone{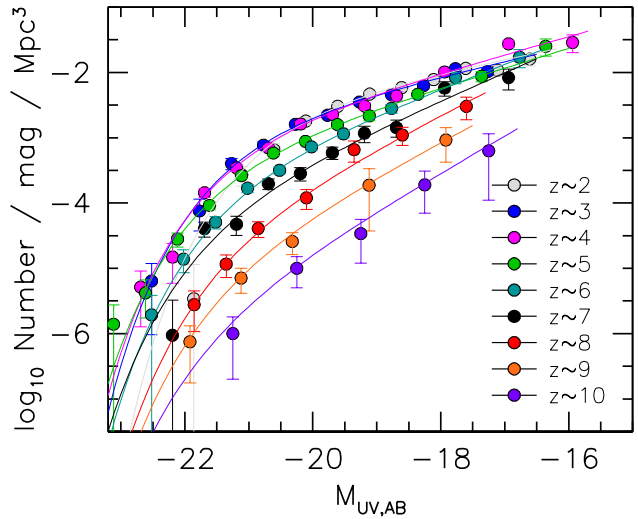}
\caption{The stepwise LF constraints (\textit{solid circles})) we
  derive on the $UV$ LFs at $z\sim2$, $z\sim3$, $z\sim4$, $z\sim5$,
  $z\sim6$, $z\sim7$, $z\sim8$, and $z\sim9$ based on our
  comprehensive blank-field searches with {\it HST} (\textit{shown in
    grey, blue, magenta, green, cyan, black, red, orange, and dark
    purple, respectively}).  The recent stepwise LF constraints at
  $z\sim10$ from Oesch et al.\ (2018a) are shown with the dark purple
  circles.  The best-fit Schechter LFs are shown with the grey, blue,
  magenta, green, cyan, black, red, orange, and dark purple lines,
  respectively.\label{fig:lfall}}
\end{figure}

In deriving LFs from our samples, we need to account for our
selections suffering from a low level of contamination from lower
redshift sources due to noise in our photometry.  Contamination is
estimated and included in a very similar way to that done in Bouwens
et al.\ (2015).  In the Bouwens et al.\ (2015) study, contamination
rates were estimated by performing degradation experiments on the
deepest {\it HST} observations.  Bona-fide high-redshift sources and
low redshift contaminants were first identified in those data.  Noise
was then added to the observations to emulate the properties of the
shallower observations, and sources were selected from these shallower
data.  The contamination rate was determined by determining which
fraction of selected sources in the shallower data were clearly at
lower redshift in the deeper data.  The typical contamination
fractions are estimated to be $\lesssim$5\% but reach contamination
fractions as high as $\sim$10\% in the faintest magnitude bin.

\begin{deluxetable*}{lclclc}
\tablewidth{0pt}
\tabletypesize{\footnotesize}
\tablecaption{Stepwise Determination of the rest-frame $UV$ LF at $z\sim4$, $z\sim5$, $z\sim6$, $z\sim7$, $z\sim8$, and $z\sim9$ using the SWML method from the HUDF, HFF parallel fields, and a comprehensive set of blank search fields.\tablenotemark{a}\label{tab:swlf}}
\tablehead{
\colhead{$M_{1600,AB}$} & \colhead{$\phi_k$ (Mpc$^{-3}$ mag$^{-1}$)} & \colhead{$M_{1600,AB}$} & \colhead{$\phi_k$ (Mpc$^{-3}$ mag$^{-1}$)} & \colhead{$M_{1600,AB}$} & \colhead{$\phi_k$ (Mpc$^{-3}$ mag$^{-1}$)}}
\startdata
\multicolumn{2}{c}{$z\sim2$ galaxies} & \multicolumn{2}{c}{$z\sim5$ galaxies} & \multicolumn{2}{c}{$z\sim8$ galaxies}\\
$-$21.86 & 0.000003$\pm$0.000008 & $-$23.11 & 0.000001$\pm$0.000001 & $-$21.85 & 0.000003$\pm$0.000002\\
$-$21.11 & 0.000270$\pm$0.000089 & $-$22.61 & 0.000004$\pm$0.000002 & $-$21.35 & 0.000012$\pm$0.000004\\
$-$20.61 & 0.000661$\pm$0.000154 & $-$22.11 & 0.000028$\pm$0.000007 & $-$20.85 & 0.000041$\pm$0.000011\\
$-$20.11 & 0.001797$\pm$0.000231 & $-$21.61 & 0.000092$\pm$0.000013 & $-$20.10 & 0.000120$\pm$0.000040\\
$-$19.61 & 0.003031$\pm$0.000301 & $-$21.11 & 0.000262$\pm$0.000024 & $-$19.35 & 0.000657$\pm$0.000233\\
$-$19.11 & 0.004661$\pm$0.000353 & $-$20.61 & 0.000584$\pm$0.000044 & $-$18.60 & 0.001100$\pm$0.000340\\
$-$18.61 & 0.005855$\pm$0.000437 & $-$20.11 & 0.000879$\pm$0.000067 & $-$17.60 & 0.003020$\pm$0.001140\\
$-$18.11 & 0.007765$\pm$0.000617 & $-$19.61 & 0.001594$\pm$0.000156 &  & \\
$-$17.61 & 0.011541$\pm$0.000835 & $-$19.11 & 0.002159$\pm$0.000346 & \multicolumn{2}{c}{$z\sim9$ galaxies}\\
$-$17.11 & 0.010795$\pm$0.002006 & $-$18.36 & 0.004620$\pm$0.000520 & $-$21.92 & 0.000001$\pm$0.000001\\
$-$16.61 & 0.015992$\pm$0.003437 & $-$17.36 & 0.008780$\pm$0.001540 & $-$21.12 & 0.000007$\pm$0.000003\\
 &  & $-$16.36 & 0.025120$\pm$0.007340 & $-$20.32 & 0.000026$\pm$0.000009\\
\multicolumn{2}{c}{$z\sim3$ galaxies} &  &  & $-$19.12 & 0.000187$\pm$0.000150\\
$-$22.52 & 0.000006$\pm$0.000005 & \multicolumn{2}{c}{$z\sim6$ galaxies} & $-$17.92 & 0.000923$\pm$0.000501\\
$-$21.77 & 0.000076$\pm$0.000038 & $-$22.52 & 0.000002$\pm$0.000002 &  & \\
$-$21.27 & 0.000402$\pm$0.000078 & $-$22.02 & 0.000014$\pm$0.000005 & \multicolumn{2}{c}{Oesch et al.\ (2018a)}\\
$-$20.77 & 0.000769$\pm$0.000117 & $-$21.52 & 0.000051$\pm$0.000011 & \multicolumn{2}{c}{$z\sim10$ galaxies}\\
$-$20.27 & 0.001607$\pm$0.000157 & $-$21.02 & 0.000169$\pm$0.000024 & $-$22.25 & $<$0.000002\\
$-$19.77 & 0.002205$\pm$0.000189 & $-$20.52 & 0.000317$\pm$0.000041 & $-$21.25 & 0.000001$\pm$0.000001\\
$-$19.27 & 0.003521$\pm$0.000239 & $-$20.02 & 0.000724$\pm$0.000087 & $-$20.25 & 0.000010$\pm$0.000005\\
$-$18.77 & 0.004557$\pm$0.000297 & $-$19.52 & 0.001147$\pm$0.000157 & $-$19.25 & 0.000034$\pm$0.000022\\
$-$18.27 & 0.006258$\pm$0.000437 & $-$18.77 & 0.002820$\pm$0.000440 & $-$18.25 & 0.000190$\pm$0.000120\\
$-$17.77 & 0.011417$\pm$0.000656 & $-$17.77 & 0.008360$\pm$0.001660 & $-$17.25 & 0.000630$\pm$0.000520\\
$-$17.27 & 0.010281$\pm$0.001368 & $-$16.77 & 0.017100$\pm$0.005260 &  & \\
 &  &  &  &  & \\
\multicolumn{2}{c}{$z\sim4$ galaxies} & \multicolumn{2}{c}{$z\sim7$ galaxies} &  & \\
$-$22.69 & 0.000005$\pm$0.000004 & $-$22.19 & 0.000001$\pm$0.000002 &  & \\
$-$22.19 & 0.000015$\pm$0.000009 & $-$21.69 & 0.000041$\pm$0.000011 &  & \\
$-$21.69 & 0.000144$\pm$0.000022 & $-$21.19 & 0.000047$\pm$0.000015 &  & \\
$-$21.19 & 0.000344$\pm$0.000038 & $-$20.69 & 0.000198$\pm$0.000036 &  & \\
$-$20.69 & 0.000698$\pm$0.000068 & $-$20.19 & 0.000283$\pm$0.000066 &  & \\
$-$20.19 & 0.001624$\pm$0.000131 & $-$19.69 & 0.000589$\pm$0.000126 &  & \\
$-$19.69 & 0.002276$\pm$0.000199 & $-$19.19 & 0.001172$\pm$0.000336 &  & \\
$-$19.19 & 0.003056$\pm$0.000388 & $-$18.69 & 0.001433$\pm$0.000419 &  & \\
$-$18.69 & 0.004371$\pm$0.000689 & $-$17.94 & 0.005760$\pm$0.001440 &  & \\
$-$17.94 & 0.010160$\pm$0.000920 & $-$16.94 & 0.008320$\pm$0.002900 &  & \\
$-$16.94 & 0.027420$\pm$0.003440 &  &  &  & \\
$-$15.94 & 0.028820$\pm$0.008740 &  &  &  & 
\enddata
\tablenotetext{a}{These binned stepwise LF parameters represent updates to those derived in Bouwens et al.\ (2015).}
\end{deluxetable*}

\begin{deluxetable}{ccccc}
\tablewidth{0pt}
\tabletypesize{\footnotesize}
\tablecaption{
Determinations of the Schechter Parameters for the rest-frame $UV$ LFs
at $z\sim4$, $z\sim5$, $z\sim6$, $z\sim7$, $z\sim8$, and $z\sim9$ from the HUDF, HFF parallels, and a comprehensive set of other blank search fields\tablenotemark{a}\label{tab:lfparm}}
\tablehead{
\colhead{Dropout} & \colhead{} & \colhead{} & \colhead{$\phi^*$ $(10^{-3}$} & \colhead{} \\
\colhead{Sample} & \colhead{$<z>$} &
\colhead{$M_{UV} ^{*}$} & \colhead{Mpc$^{-3}$)} & \colhead{$\alpha$}}
\startdata
$U_{275}$ & 2.1 & $-$20.28$\pm$0.09 & 4.0$_{-0.4}^{+0.5}$ & $-$1.52$\pm$0.03\\
$U_{336}$ & 2.9 & $-$20.87$\pm$0.09 & 2.1$_{-0.3}^{+0.3}$ & $-$1.61$\pm$0.03\\
$B$ & 3.8 & $-$20.93$\pm$0.08 & 1.69$_{-0.20}^{+0.22}$ & $-$1.69$\pm$0.03\\ 
$V$ & 4.9 & $-$21.10$\pm$0.11 & 0.79$_{-0.13}^{+0.16}$ & $-$1.74$\pm$0.06\\
$i$ & 5.9 & $-$20.93$\pm$0.09 & 0.51$_{-0.10}^{+0.12} $ & $-$1.93$\pm$0.08\\
$z$ & 6.8 & $-$21.15$\pm$0.13 & 0.19$_{-0.06}^{+0.08} $ & $-$2.06$\pm$0.11\\
$Y$ & 7.9 & $-$20.93$\pm$0.28 & 0.09$_{-0.05}^{+0.09} $ & $-$2.23$\pm$0.20\\
$J$ & 8.9 & $-$21.15 (fixed) & 0.021$_{-0.009}^{+0.014}$ & $-$2.33$\pm$0.19\\\\
\multicolumn{5}{c}{Oesch et al.\ (2018a)}\\
$J$ & 10.2 & $-$21.19 (fixed) & 0.0042$_{-0.0022}^{+0.0045}$ & $-$2.38$\pm$0.28\\
\enddata
\tablenotetext{a}{These Schechter parameters represent updates to those derived in Bouwens et al.\ (2015)
and incorporate all the new search results indicated in Table~\ref{tab:sampnumbers}.}
\end{deluxetable}

In deriving constraints on the $UV$ LF from a comprehensive set of
search fields, we rely on the same sample of sources that Bouwens et
al.\ (2015) utilize over all fields, while also including constraints
from the new data sets.  Combining the new samples with the $z\sim
2$-9 samples from Bouwens et al.\ (2015) and Bouwens et al.\ (2016),
our new analysis contains contains 5766, 6332, 7240, 3449, 1066, 601,
246, and 33 sources at $z\sim2$, $z\sim3$, $z\sim4$, $z\sim5$,
$z\sim6$, $z\sim7$, $z\sim8$, and $z\sim9$, respectively.

\begin{figure*}
\epsscale{1.15}
\plotone{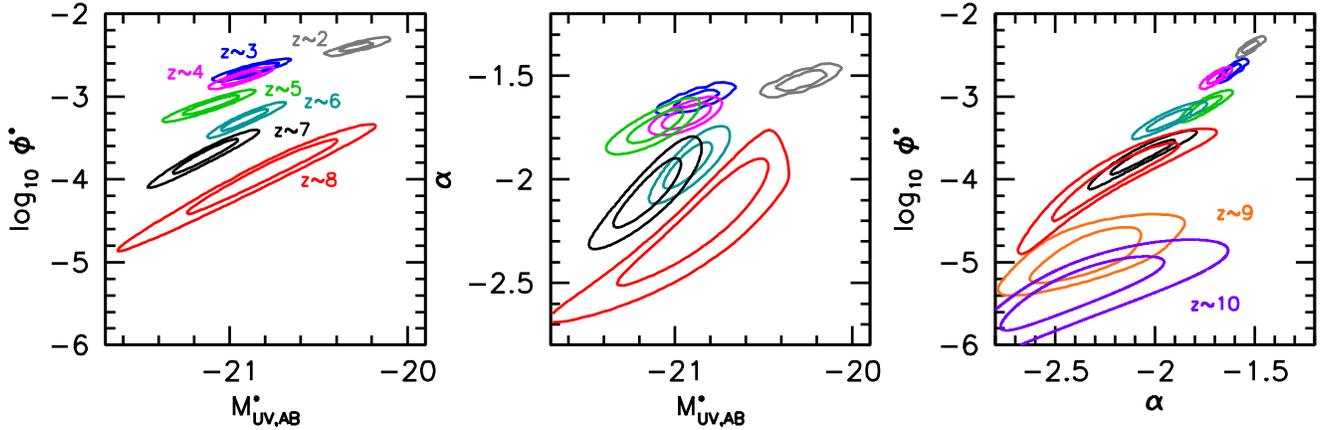}
\caption{68\% and 95\% confidence intervals on various pairs of
  parameters in a Schechter representation of the $UV$ LF at $z\sim2$,
  $z\sim3$, $z\sim4$, $z\sim5$, $z\sim6$, $z\sim7$, $z\sim8$,
  $z\sim9$, and $z\sim10$.  Given our relatively poor constraints on
  the bright end form of the $z\sim9$ and $z\sim10$ LFs and thus
  $M^*$, no confidence intervals are presented in the left and center
  panels for the LF results at $z\sim9$ and $z\sim10$.  The $z\sim10$
  constraints are based on the Oesch et al.\ (2018a) analysis.  The
  normalization $\phi^*$ of the $UV$ LF and the faint-end slope
  $\alpha$ smoothly increase and flatten from $z\sim9$ to $z\sim2$,
  while the characteristic luminosity $M^*$ shows no substantial
  evolution from $z\sim8$ to $z\sim3$.\label{fig:contours}}
\end{figure*}

In deriving our new LF constraints, we adopt the same approach as we
describe in Bouwens et al.\ (2015) where we find the binned LF which
maximizes the likelihood ${\cal L}$ of matching the binned number
counts in all of our fields
\begin{equation}
{\cal L}=\Pi_{field} \Pi_i p(m_i)
\label{eq:ml}
\end{equation}
where $i$ runs over all magnitude intervals in each of our search
fields.  For our $z\sim2$-8 samples, we take the probability $p(m_i)$
to be 
\begin{equation}
p(m_i) = \left (\frac{n_{expected,i}}{\Sigma_j n_{expected,j}} \right )^{n_{observed,i}}
\end{equation}
for all sources in our $z=2$-8 samples, where $n_{expected,i}$ and
$n_{expected,j}$ the expected number of sources in magnitude intervals
$i$ and $j$ and $n_{observed,i}$ is the observed number of sources in
magnitude interval $i$.  As such, our $z=2$-8 LFs are computed using
the standard stepwise maximum likelihood procedure (Efstathiou et
al.\ 1988) to take advantage of the modest number of sources found in
each search field and overcome large-scale structure uncertainties.

Given the much smaller number of sources that are available per search
field to determine the shape of the $UV$ LF for our $z\sim9$ samples,
we compute the probabilities in this redshift range assuming that the
counts are Poissonian distributed:
\begin{equation}
p(m_i) = \Pi_{j} e^{-n_{expected,j}} \frac{(n_{expected,j})^{n_{observed,j}}}{(n_{observed,j})!}
\label{eq:ml}
\end{equation}
For our stepwise LFs, we generally adopt a width of 0.5-mag for our
$z=2$-8 and 0.8-mag for our LFs at $z=9$-10.

We compute the expected number of sources in a given magnitude
interval $n_{expected,i}$ as
\begin{equation}
n_{expected,i} = \Sigma _{j} \phi_j V_{i,j}
\label{eq:numcountg}
\end{equation}
where $V_{i,j}$ is the effective volume over which a source of
absolute magnitude $j$ might be expected to be found in the observed
magnitude interval $i$.  The effective volume $V_{i,j}$ is computed
from extensive Monte-Carlo simulations where we add artificial sources
of absolute magnitude $j$ to the real observations and then quantify
the fraction of these sources that will be both selected as part of a
given high-redshift samples and measured to have an apparent magnitude
$i$.

In deriving $n_{observed,i}$ from our large $z\sim2$, 3, 4, 5, 6, 7,
8, and 9 selections, we use the measured total magnitude of sources in
the $V_{606}$, $I_{814}$, $i_{775}$, $z_{850}$, $Y_{105}$, $J_{125}$,
$H_{160}$, and $H_{160}$, respectively, since those magnitudes lie
closest to rest-frame 1600\AA.  For some search fields and redshift
samples, flux measurements are not available in these bands.  For our
HFF selections, magnitude measurements in the $I_{814}$, $Y_{105}$,
and $Y_{105}$ bands, respectively, are used for our $z\sim4$,
$z\sim5$, and $z\sim6$ selections.  For the wide CANDELS fields, flux
measurements in the $J_{125}$ band are used for our $z\sim5$,
$z\sim6$, and $z\sim7$ selections.  For $z\sim3$ sources over the
UVUDF, flux measurements in the $i_{775}$ band are used.

\begin{figure*}
\epsscale{1.15}
\plotone{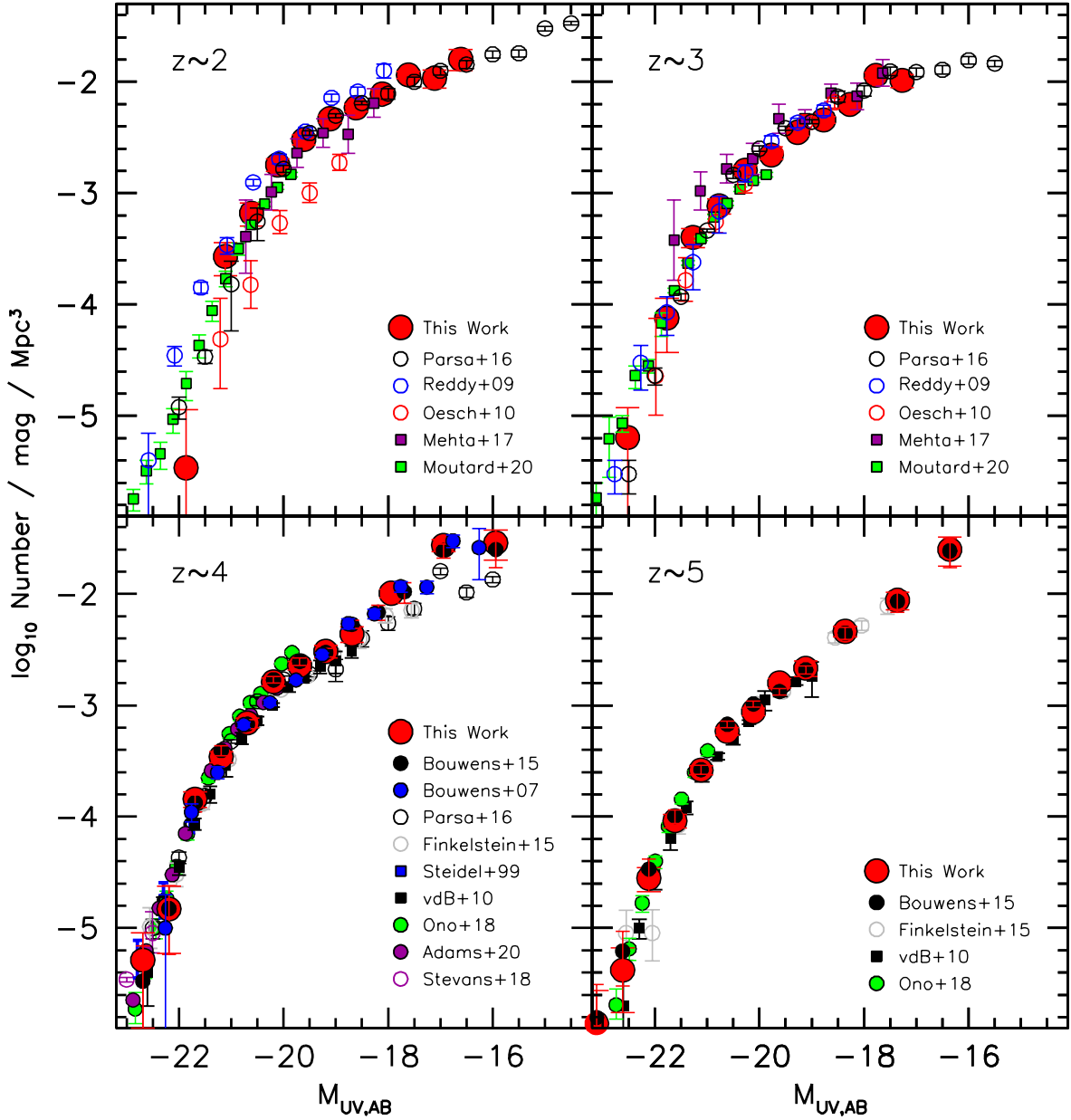}
\caption{Comparison of the new $z=2$-5 $UV$ LFs we derived updating
  and extending the redshift baseline of the Bouwens et al.\ (2015)
  and Bouwens et al.\ (2016: \textit{solid red circles}) results
  against previous blank-field LF results in the literature including
  those from Reddy \& Steidel (2009: \textit{open blue circles}),
  Oesch et al.\ (2010: \textit{open red circles}), Mehta et
  al.\ (2017: \textit{magenta solid squares}), Parsa et al.\ (2016:
  \textit{black open circles}), Moutard et al.\ (2020: \textit{solid
    green squares}), Finkelstein et al.\ (2015: \textit{open gray
    circles}), van den Bosch et al.\ (2010: \textit{black solid
    squares}), Steidel et al.\ (1999: \textit{blue solid squares}),
  Ono et al.\ (2018: \textit{solid green circles}), Adams et
  al.\ (2018: \textit{solid violet circles}), Stevans et al.\ (2018:
  \textit{open violet circles}), and Bouwens et al.\ (2007:
  \textit{solid blue circles}).  The present determinations are in
  broad agreement with previous work.  No results from lensing cluster
  studies are included here to keep the discussion simple, but will be
  included in a forthcoming companion paper focusing on the HFF
  clusters.\label{fig:comp25}}
\end{figure*}

\begin{figure*}
\epsscale{1.15}
\plotone{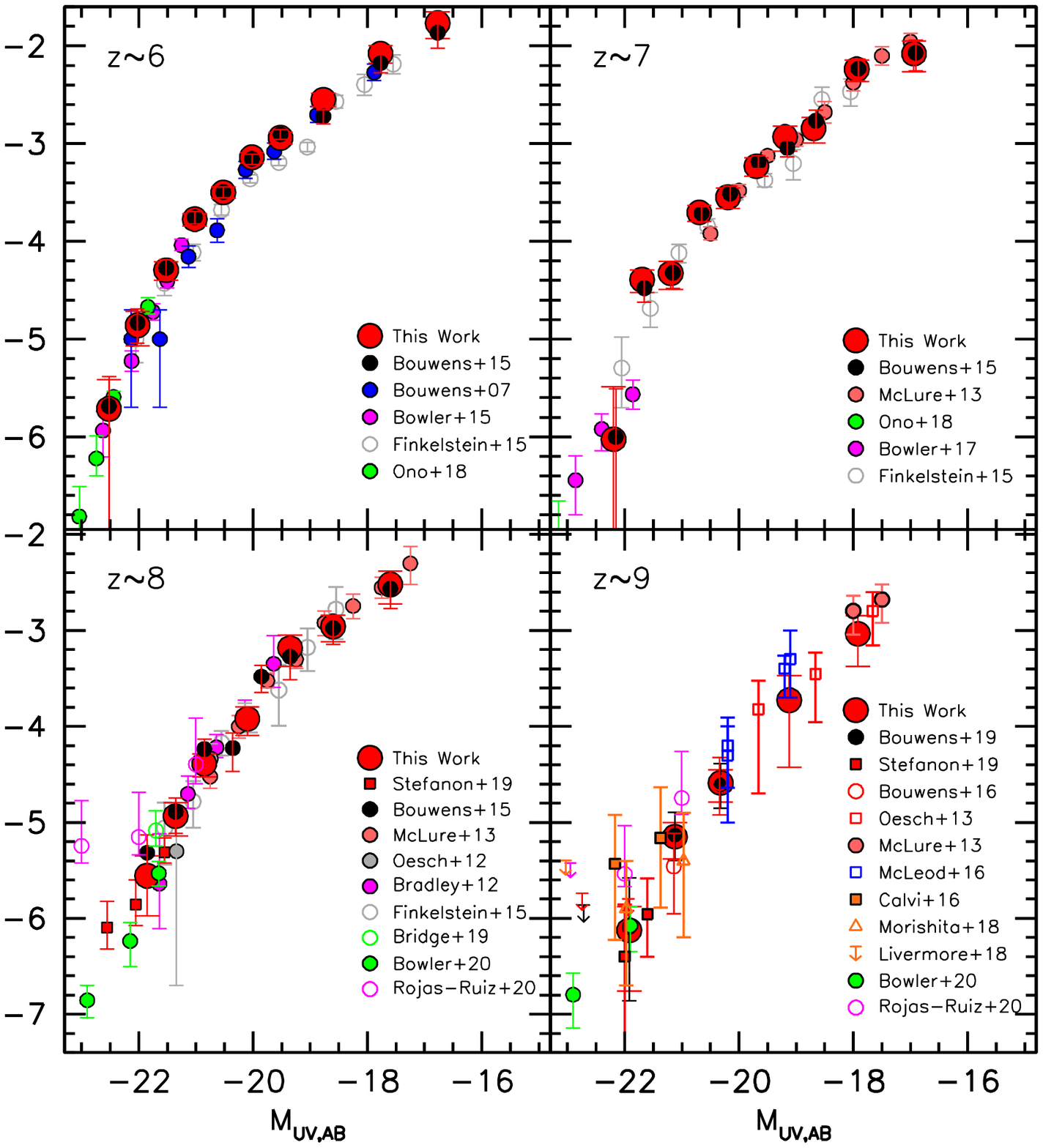}
\caption{Similar to Figure~\ref{fig:comp25} but for our newly derived
  LFs at $z=6$-9.  Included in these comparisons are the results of
  Bouwens et al.\ (2007: \textit{solid blue circles}), Bouwens et
  al.\ (2015: \textit{solid black circles}), Ono et al.\ (2018:
  \textit{solid green circles}), Bowler et al.\ (2015: \textit{solid
    magenta circles}), Finkelstein et al.\ (2015: \textit{open gray
    circles}), McLure et al.\ (2013: \textit{solid light red
    circles}), Bowler et al.\ (2017: \textit{solid magenta circles}),
  Oesch et al.\ (2012: \textit{solid gray circles}), Bradley et
  al.\ (2012: \textit{solid magenta circles}), Bridge et al.\ (2019:
  \textit{open green circle}), Bowler et al.\ (2020: \textit{green
    solid circles}), Rojas-Ruiz et al.\ (2020: \textit{open magenta
    circles}), Bouwens et al.\ (2019: \textit{solid black squares}),
  Stefanon et al.\ (2019: \textit{solid red squares}), Bouwens et
  al.\ (2016: \textit{open red circles}), Oesch et al.\ (2013:
  \textit{open red squares}), McLeod et al.\ (2016: \textit{open blue
    squares}), Calvi et al.\ (2016: \textit{solid light red squares}),
  Morishita et al.\ (2018: \textit{open red triangle}), and Livermore
  et al.\ (2018: \textit{light red upper limits}).\label{fig:comp69}}
\end{figure*}

In making use of the search constraints to derive LF results, we only
consider results to specific limiting magnitudes to avoid having the
results be significantly impacted by uncertain completeness
corrections or contamination rates.  We adopt the same limiting
magnitudes as Bouwens et al.\ (2015), except in the cases of the new
samples considered here, including our $z\sim2$-3 samples from the
ERS, HDUV, and UVUDF data sets where 26.5 mag, 28.0 mag, and 29.0 mag,
respectively, our $z\sim4$-9 HFF parallel samples where 29.0 mag
limits are used, and our new $z\sim9$ XDF, HUDF09-1, and HUDF09-2
samples where 30.0, 29.0, and 29.0 mag, respectively, are used.

Finally, over the HFF parallel fields, we accounted for the estimated
magnification factors using the lensing models from Merten (2016).
The approximate lensing magnification that we applied in magnitudes
for each parallel field is provided in Table~\ref{tab:hffpar_mu} for
sources at $z\sim6$.  The magnification factors at other redshifts are
very similar.  The search volumes and luminosities were reduced and
scaled according to the lensing magnification in each field.

We present updated stepwise determinations of the $UV$ LF at $z\sim2$,
$z\sim3$, $z\sim4$, $z\sim5$, $z\sim6$, $z\sim7$, $z\sim8$, and
$z\sim9$ in Figure~\ref{fig:lfall} and Table~\ref{tab:swlf}, along
with the new $z\sim10$ results from Oesch et al.\ (2018a) designed to
complement this study.  $UV$ LF results are similarly derived using a
Schechter parameterization by first fitting for the shape of the LF as
in the SWML approach as in Sandage et al.\ (1979) and then determining
the normalization $\phi^*$.  The best-fit Schechter results are
provided in Table~\ref{tab:lfparm}, together with the $z\sim10$
results of Oesch et al.\ (2018a).  For our $z\sim9$ LF determinations,
we fix the characteristic luminosity $M^*$ to the value implied by the
fitting formula derived in \S4.2, i.e., $-$21.15 mag.  For the Oesch
et al.\ (2018a) $z\sim10$ LF constraints, we similarly fixed $M^*$ to
be $-21.19$ mag, while fitting for constraints on $\phi^*$ and
$\alpha$.

Included in our best-fit LFs are the quoted stepwise constraints from
a large variety of different ground-based probes including Stefanon et
al.\ (2019), Bowler et al.\ (2015), and the brightest two magnitude
bins in Bowler et al.\ (2015) where their selection of bright $z\sim7$
galaxies should be the most complete.

Finally, the new constraints on the UV LF at $z\sim6$ and $z\sim7$
faintward of $-$23 mag from Ono et al.\ (2018) are included in our
fits.  If constraints brighter than $-$23 mag are included in our $UV$
LF fits, we find that the LF constraints are not well represented by a
Schechter function-type form and the characteristic luminosity is
driven towards higher values.
  
Figure~\ref{fig:contours} shows the 68\% and 95\% confidence intervals
we compute for various two-dimentional projections of the Schechter
parameters.  We discuss evolution in the Schechter parameters in
\S\ref{sec:evolsch}.

\section{Discussion}

\subsection{Comparison with Previous LF Results}

There is now a quite substantial body of work on the $UV$ LF at high
redshift, from $z\sim2$-3 (Madau et al.\ 1996; Steidel et al.\ 1999;
Reddy \& Steidel 2009; Oesch et al.\ 2010; Alavi et al.\ 2016) to
$z\sim4$-5 (e.g., Bouwens et al.\ 2007; van der Berg et al.\ 2010;
Bouwens et al.\ 2015; Parsa et al.\ 2016) to $z\sim6$-10 (Bouwens et
al.\ 2008; Oesch et al.\ 2010, 2012; McLure et al.\ 2013; Schenker et
al.\ 2013; Bouwens et al.\ 2015, 2016, 2017; Finkelstein et al.\ 2015;
Oesch et al.\ 2015; McLeod et al.\ 2016; Livermore et al.\ 2018; Atek
et al.\ 2018).

It is useful to compare the present determinations of the $UV$ LFs
against many previous determinations to quantify possible differences
in the results.  Given that the present results utilize blank-field
surveys to arrive at the LFs results, we focus on comparisons with
previous blank-field determinations to keep the comparisons most
direct.

Accordingly, in Figures~\ref{fig:comp25} and \ref{fig:comp69}, we
provide a comprehensive set of comparisons of our new $z=2$-9 LF
results from the HFFs with a variety of noteworthy previous work,
including Steidel et al.\ (1999), Bouwens et al.\ (2007), Reddy \&
Steidel (2009), Oesch et al.\ (2010), van der Berg et al.\ (2010),
Bradley et al.\ (2012), Oesch et al.\ (2012), McLure et al.\ (2013),
Bouwens et al.\ (2015), Bowler et al.\ (2015), Finkelstein et
al.\ (2015), Bouwens et al.\ (2016), Parsa et al.\ (2016), Bouwens et
al.\ (2016), McLeod et al.\ (2016), Ono et al.\ (2018), and Stefanon
et al.\ (2019).

We consider the redshift intervals in turn, below:\\\vskip-0.1cm

\noindent $z\sim2$-3: For $UV$ luminosities of $\sim$0.1 $L^*$ ($-$20
mag to $-$17 mag), most existing LF results at $z\sim2$ and $z\sim3$
are broadly in agreement.  This is especially true brightward of
$-$20, where essentially all recent studies (this study; Reddy \&
Steidel 2009; Oesch et al.\ 2010; Parsa et al.\ 2016; Mehta et
al.\ 2017; Moutard et al.\ 2020) show approximately (modulo $<$0.2-mag
differences) the same bright-end cut-off.  In contrast to the $z\sim3$
results, the absolute magnitude of the cut-off at $z\sim2$ varies much
more substantially, occurring$\sim$0.7 mag brighter in the Reddy \&
Steidel (2009) case than in the Oesch et al.\ (2010)
case.

The only apparent exception to this are the $z\sim2$ results of Oesch
et al.\ (2010), which appear to be a factor of $\sim3$ lower than the
other $z\sim2$ LF results.  To investigate this difference, we
constructed a $z\sim1.9$ sample of galaxies using the same
$U_{275}$-dropout criteria as given in Oesch et al.\ (2010) and
compared it to the present selection of $z\sim2$ galaxies to the same
25.5-mag limit in the $B_{435}$ band.  Our $z\sim2$ selection shows
$\sim$2.5$\times$ more sources, i.e., 245, to the same magnitude limit
as Oesch et al.\ (2010) use.  If the estimate of the selection volume
at $z\sim2$ in these previous studies is similar to the present
estimate, this would largely explain the difference in our LF results.
While the Oesch et al.\ (2010) results seem very reasonable in
isolation, the estimated selection volume in $z\sim2$ samples is very
sensitive to the expected S/N in the $U_{275}$ and $U_{336}$ bands,
which in turn is sensitive to the source size and surface brightness.
Additionally, a difference in the mean redshift of the Oesch et
al.\ (2010) $z\sim1.9$ election and the present selection $z\sim2.1$
(typical redshift uncertainties for sources is $\Delta z\sim0.2$-0.3)
likely contribute to the observed differences.\\\vskip-0.1cm

\noindent $z\sim4$-5: For comparisons between our new $z\sim4$ and
$z\sim5$ results and previous determinations, we note good agreement
between our new $z\sim4$ and $z\sim5$ LF determinations and various
comparison luminosity functions from the literature
(Figure~\ref{fig:comp25}: Steidel et al.\ 1999; van der Berg et
al.\ 2010; Bouwens et al.\ 2015; Finkelstein et al.\ 2015; Parsa et
al.\ 2016; Ono et al.\ 2018; Stevans et al.\ 2018; Adams et al.\ 2020)
at the high luminosity end.  At lower luminosities, i.e., reaching to
$-17$ to $-16$ mag, our $z\sim4$ LF determinations are in good
agreement with our previous determinations (Bouwens et al.\ 2007,
Bouwens et al.\ 2015) but a factor of 1.5-2 higher than those in
Finkelstein et al.\ (2015) and Parsa et al.\ (2016).  One potential
explanation for the difference could be Finkelstein et al.\ (2015) and
Parsa et al.\ (2016)'s use of a $1/V_{max}$ estimator to derive the
Schechter function parameters.  LF determinations using the
$1/V_{max}$ estimator can be impacted if the search fields probing a
particular luminosity range show a significant overdensity or
underdensity of sources. In the case of the HUDF/XDF, our best-fit
$z\sim4$ LF determination suggests we would find 40$\pm$7\% more
$z\sim4$ sources in the HUDF/XDF data than what we actually find,
suggesting that the HUDF/XDF region may be underdense by $\sim$30\%.
If the Finkelstein et al.\ (2015)/Parsa et al.\ (2016) determinations
are impacted by this issue, it could result in $\Delta\alpha \sim
-0.1$ shallower values for the faint-end slope $\alpha$, consistent
with the observed differences.\\\vskip-0.1cm

\noindent $z\sim6$-7: As at $z\sim4$-5, our new constraints on the
$UV$ LFs at $z\sim6$-7 are in broad agreement
(Figure~\ref{fig:comp25}) with previous determinations, e.g., Bouwens
et al.\ (2007), McLure et al.\ (2013), Bouwens et al.\ (2015), Bowler
et al.\ (2015), Ono et al.\ (2018), and Finkelstein et al.\ (2015).
At intermediate luminosities, i.e., $-19$ mag, where the results would
be sensitive to the faintest sources in the CANDELS selections and the
estimated selection volumes, the Finkelstein et al.\ (2015) $z\sim6$
LF results (and to lesser extent their $z\sim7$ results) are a factor
of $\sim$2 lower than our new and previous LF results.  If the
selection volumes in this regime were overestimated due to reliance on
the \textit{selected} population of $z\sim6$ galaxies \textit{from
  CANDELS} (which would tend to include only the highest surface
brightness sources) for the completeness estimates, this could explain
the differences at $\sim$$-$19 mag.  In any case, at $z\sim6$-7, we
consistently recover the same volume density of sources at $-19$ mag
regardless of whether we rely on the significantly deeper HFF or
CANDELS data.\\\vskip-0.1cm

\noindent $z\sim8$-9: Our new results at $z\sim8$-9 are in excellent
agreement with essentially all of the latest determinations at these
redshifts (Oesch et al.\ 2012, 2013; Bradley et al.\ 2012; McLure et
al.\ 2013; Bouwens et al.\ 2015, 2016, 2019; Finkelstein et al.\ 2015;
Calvi et al.\ 2016; McLeod et al.\ 2016; Morashita et al.\ 2018;
Livermore et al.\ 2018; Bridge et al.\ 2019; Bowler et al.\ 2020;
Rojas-Ruiz et al.\ 2020).  At the bright end, the new $z\sim8$-9 LF
constraints from Stefanon et al.\ (2019: see also Stefanon et
al.\ 2017b) and Bowler et al.\ (2020) from very wide-area ($\sim2$-6
deg$^2$) searches are consistent with what we derive, but extend to
higher luminosities.  While the Calvi et al.\ (2016) results from the
BoRG/HIPPIES pure-parallel fields are somewhat in excess of our own,
this is without inclusion of the {\it Spitzer}/IRAC observations into
the analysis to exclude lower-redshift interlopers and AGN.  The
Morishita et al.\ (2018) analyses of the BoRG/HIPPIES fields are in
much better agreement with our results, supporting this conclusion.
The $z\sim8$ LF from Rojas-Ruiz et al.\ (2018) at $-$23 mag is clearly
higher than the other LF determinations that probe this regime
(Stefanon et al.\ 2019; Bowler et al.\ 2020), but is based on only a
single source and therefore the uncertainties are large.  At fainter
luminosities, the faint-end results from McLure et al.\ (2018) and
McLeod et al.\ (2016) are also encouragingly consistent with the new
LF results we have obtained including all six parallel fields in the
HFF program.

\begin{figure}
\epsscale{1.17}
\plotone{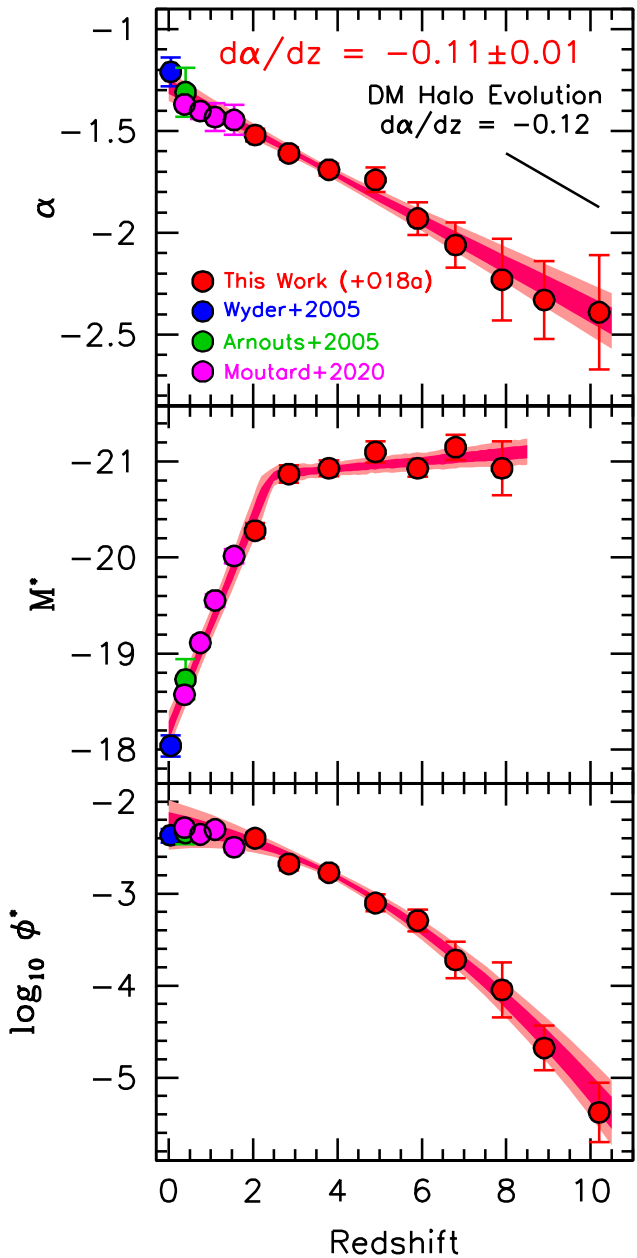}
\caption{Determinations of the faint-end slope $\alpha$,
  characteristic luminosity $M^*$, and normalization $\phi^*$ to the
  $UV$ LF derived at $z=2$-10 in this work and Oesch et al.\ (2018a:
  O18a) from blank-field observations alone (\textit{solid red
    circles}).  The plotted $z\sim2$-3 determinations are our new
  results derived using the HDUV+UVUDF+ERS WFC3/UVIS data sets, while
  the plotted $z\sim4$-9 results represent updates to our earlier
  determinations from Bouwens et al.\ (2015, 2016) including new
  constraints from the HFF-parallel fields.  The Wyder et al.\ (2005)
  faint-end slope determination at $z\sim0.055$ is shown, an average
  of the $z\sim0.2$-0.4 determinations by Arnouts et al.\ (2005), and
  the $z=0.3$-0.45, $z=0.6$-0.9, $z=0.9$-1.3, and $z=1.3$-1.8
  determinations of Moutard et al.\ (2020).  The light and dark
  reddish magenta contours show our 68\% and 95\% constraints on the
  evolution in the faint-end slope $\alpha$, $M^*$, and $\log_{10}
  \phi^*$ inferred from a fit to the present LF results.  The plotted
  $z<3$ constraints on $M^*$ from Wyder et al.\ (2005), Arnouts et
  al.\ (2005), and Moutard et al.\ (2020) are used in deriving the
  best-fit relations.  The present fits represent an update to the
  determinations in Parsa et al.\ (2016) who look at the evolution of
  the LF parameters over a similar redshift baseline (see also Moutard
  et al.\ 2020, Bowler et al.\ 2020, and Finkelstein 2016).
  Interestingly, the observed evolution can be remarkably well
  explained by the predicted evolution in the halo mass function and a
  fixed star-formation efficiency model (see
  \S\ref{sec:evolsch}).\label{fig:alpha}}
\end{figure}

\subsection{Evolution in $\alpha$, $M^*$, and $\phi^*$\label{sec:evolsch}}

As in our previous comprehensive analyses of the $UV$ LF at
$z\sim4$-10 based on blank-field observations (Bouwens et al.\ 2015),
we can use our improved constraints on the $UV$ LF to examine
evolution in the Schechter parameters.

While the evolution in these quantities is already clear based on
previous work (e.g., Bouwens et al.\ 2015; Bowler et al.\ 2015;
Finkelstein et al.\ 2015), the new observations allow us to improve
our previous determinations even further to map out the evolutionary
trends.  While recognizing the value of $UV$ LF results that rely on
lensing magnification by galaxy clusters, we intentionally do not
include them in the present determinations to avoid any systematics
that might result from managing uncertainties in the lensing models or
uncertainties in the sizes of the lowest luminosity sources.

While our primary interest here is in looking at the faint-end slope
trend, we will also look at how the other two Schechter parameters
evolve.  Based on the plotted contours in Figure~\ref{fig:contours},
the normalization $\phi^*$ shows a similarly smooth increase with
cosmic time, while the faint-end slope $\alpha$ shows a smooth
evolution from very steep values to shallower values at later points
in cosmic time.

As in our previous work, e.g., Bouwens et al.\ (2008), we assume that
the evolution is linear in $\alpha$ and $M^*$, but will take the
evolution in $\log_{10} \phi^*$ to be quadratic in form.  To account
for impact of quenching on the high star formation end of the main
sequence (e.g., Ilbert et al.\ 2013; Muzzin et al.\ 2013), we allow
for a break in the linear evolution of the $UV$ LF at $z\lesssim 2.5$,
fitting separately for the transition redshift $z_t$ and linear trend
at $z\lesssim 2.5$.  In fitting for the trend in the characteristic
luminosity, we make use of the Wyder et al.\ (2005) UV LF results at
$z\sim0.055$, the Arnouts et al.\ (2005) results at $z\sim0.3$, and
the Moutard et al.\ (2020) results over the redshift range
$z\sim0.3$-1.8.

The best-fit evolution we derive based on our blank-field LF results
at $z\sim2$-10 is the following:
\begin{eqnarray*}
M_{UV} ^{*} =& \left\{\begin{array}{ll}
               (-20.89\pm0.12) + & \textrm{for}~ z < z_t\\
               (-1.09\pm0.07) (z - z_t), \\
               (-21.03\pm0.04) + & \textrm{for}~ z > z_t\\
               (-0.04\pm0.02) (z - 6), 
               \end{array} \right. \\
\phi^* =& (0.40\pm0.04)(10^{-3} \textrm{Mpc}^{-3})~~~~~~~~~~~\\
        & ~~10^{(-0.33\pm0.02)(z-6)+(-0.024\pm0.006)(z-6)^2}\\
\alpha =& (-1.94\pm0.03) + (-0.11\pm0.01)(z-6)
\end{eqnarray*}
where $z_t = 2.46\pm0.11$.  A comparison of the best-fit trends with
the derived Schechter parameters for the $UV$ LF from $z\sim2$ to
$z\sim10$ is presented in Figure~\ref{fig:alpha}.

As in previous work, the faint-end slope $\alpha$ of the $UV$ LF is
well described by a linear flattening in $\alpha$ from
$\alpha\sim-2.3$ at $z\sim8$-10 to significantly shallower slopes,
i.e., $\alpha\sim-1.5$, at $z\sim 2$.  Moreover, an extrapolation of
our results to $z\sim0$ agree very well with the results obtained by
Wyder et al.\ (2005) at $z\sim0.055$, Arnouts et al.\ (2005) at
$z\sim0.3$, and Moutard et al.\ (2020) over the redshift range
$z\sim0.3$-1.8.

The observed change in $\alpha$ with redshift we find, i.e.,
$d\alpha/dz=-0.11\pm 0.01$ is very similar with predicted flattening
based on the evolution in the halo mass function.  Bouwens et
al.\ (2015) find that $d\alpha/dz = -0.12$ purely due to a flattening
in halo mass function with cosmic time using a simple conditional
luminosity function model.  Amazingly, the faint-end slope $\alpha$
appears to maintain a roughly linear relationship with redshift down
to $z\sim0$.  At first glance, this might seem surprising given the
increasing importance of other physical processes like AGN feedback
(e.g., Scannapieco \& Oh 2004; Croton et al.\ 2006) and the potential
impact of this feedback on star formation in lower mass halos.  The
trends we find here are very similar to what we reported in our
earlier LF study (Bouwens et al.\ 2015), i.e., $d\alpha/dz =
-0.10\pm0.03$, and also similar to the $d\alpha/dz \sim -0.11$ trend
Parsa et al.\ (2016) and Finkelstein (2016) find fitting the
then-available LF constraints in the literature.

\begin{figure}
\epsscale{1.17}
\plotone{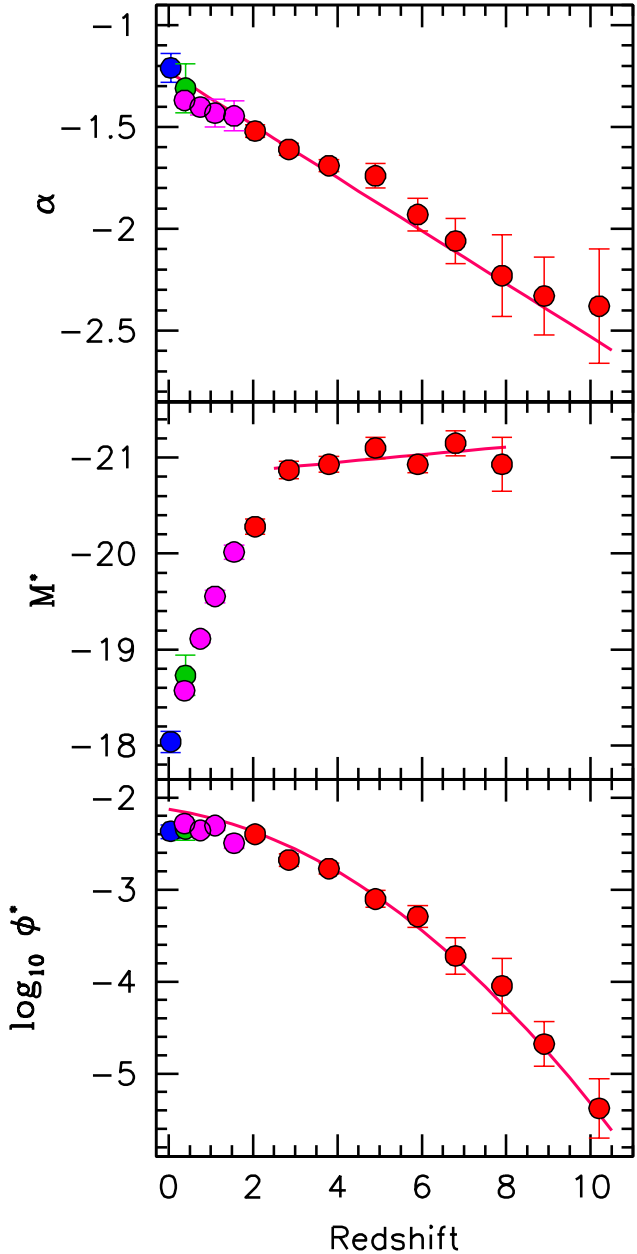}
\caption{Comparison of the observational constraints presented in
  Figure~\ref{fig:alpha} with the predictions (\textit{red lines}) of
  the simple constant star formation efficiency model presented in
  Appendix I of Bouwens et al.\ (2015), while keeping the
  characteristic luminosity $M^*$ fixed to the $-21.03-0.04(z-6)$ mag
  parameterization derived from our empirical fits
  (\S\ref{sec:evolsch}).  It is striking how observational results
  agree with the dependence predicted using the evolution of the halo
  mass function and a simple constant star formation efficiency
  model.\label{fig:modelfit}}
\end{figure}

The characteristic luminosity $M^*$ maintains a relatively fixed value
of $-21.02$ mag over the redshift range $z\sim8$ to $z\sim3$.  The
best-fit dependence of $M^*$ on redshift is $-0.04\pm0.02$ and
nominally significant at $2\sigma$.  As has been argued by Bouwens et
al.\ (2009) and Reddy et al.\ (2010), the observed exponential cut-off
at the bright end of the $UV$ LF likely occurs due to the impact of
dust extinction in sources with the highest masses and SFRs.  Galaxies
with masses and SFRs higher than some critical value (e.g., Spitler et
al.\ 2014; Stefanon et al.\ 2017b) tend to suffer sufficient
attenuation that these sources actually become fainter in the
rest-$UV$ than lower mass, lower SFR sources.  The critical $UV$
luminosity where the UV luminosity vs. SFR relationship transitions
from being positively correlated to negative correlatively appears to
set the value of the characteristic luminosity (e.g., Bouwens et
al.\ 2009; Reddy et al.\ 2010).  The relatively minimal evolution in
the characteristic luminosity $M^*$ with redshift suggests that this
critical SFR does not evolve dramatically with redshift, as Bouwens et
al.\ (2015) illustrate with the conditional luminosity function model
they present in their \S5.5.2 and Figure 20.

The normalization $\phi^*$ of the $UV$ LF increases monotonically with
cosmic time from $z\sim10$ to $z\sim2$, with a steeper dependence on
redshift from $z\sim10$ to $z\sim7$ than from $z\sim7$ to $z\sim2$.
We found that the dependence of $\log_{10} \phi^*$ with redshift could
be well described by a second-order polynomial.  The amplitude of the
second-order term, i.e., $-0.024\pm0.006$, is significant at
4$\sigma$.  The change in the dependence of $\log_{10} \phi^*$ with
redshift has been previous framed as ``accelerated'' evolution by
Oesch et al.\ (2012).  Analyses of subsequent observations in Oesch et
al.\ (2014), Oesch et al.\ (2018a), and Ishigaki et al.\ (2018: but
see also McLeod et al.\ 2016) provide further evidence for this
result.  

The observed evolution can fairly naturally be explained, using a
constant star formation efficiency model, by the evolution of the halo
mass function (e.g., Bouwens et al.\ 2008; Tacchella et al.\ 2013;
Bouwens et al.\ 2015; Mason et al.\ 2015; Oesch et al.\ 2018; Harikane
et al.\ 2018; Tacchella et al.\ 2018).  Oesch et al.\ (2018), for
example, showed with such a model that one could reproduce the
observed evolution in the dust-corrected UV luminosity from $z\sim10$
to $z\sim4$.  Adopting the conditional LF model from Bouwens et
al.\ (2015: their Appendix I), fixing the characteristic luminosity
$M^*$ to $\sim -21.03$ mag preferred from our empirical fitting
formula, and fitting for $\phi^*$ and $\alpha$, we find a best-fit
parameterization for $\phi^*$ of (0.00036 Mpc$^{-3}$)
$10^{-0.34(z-6)-0.024 (z-6)^2}$, remarkably similar to the
coefficients we recovered in deriving the LF fitting formula from the
observations.  In deriving Schechter parameters from the model LF
results from Bouwens et al.\ (2015), we minimized the square
logarithmic difference between the condition LF predictions and the
Schechter function fits over the range $-22$ to $-15$.

Fixing the characteristic luminosity $M^*$ instead to that found in
our fitting formula, i.e., $-21.03-0.04(z-6)$ mag, we find a best-fit
parameterization for $\phi^*$ of (0.00036 Mpc$^{-3}$)
$10^{-0.37(z-6)-0.025 (z-6)^2}$.  This confirms that constant star
formation efficiency models do clearly predict a second-order
dependence in the Schechter parameters, i.e., ``acceleration,''
vs. redshift.  In Figure~\ref{fig:modelfit}, we compare the
predictions of this simple model with the observational results, and
it is striking how well the evolutionary trends of such a model agrees
with the observations.  This strongly suggests that much of the
evolution of the $UV$ LF can be explained by largely explained by the
evolution of the halo mass function and an unevolved star formation
efficiency.

Given our reliance on what is currently the largest {\it HST} sample
of $z=2$-10 galaxy candidates to date, our derived evolutionary trends
arguably represent the most accurate determinations obtained to date.

\section{Summary}

In this paper, we make use of a suite of new data sets to
significantly expand current {\it HST} samples of $z\sim2$-9 galaxies
and to improve recent determinations of the $UV$ LF based on {\it HST}
data.

For our $z\sim2$-3 selections, the most important new data set are the
HDUV observations in the F275W and F336W bands over a $\sim$94
arcmin$^2$ area at $\sim$0.28$\mu$m and 0.34$\mu$m.  By combining this
data set with the $\sim$50 arcmin$^2$ WFC3/UVIS ERS (Windhorst et
al.\ 2011) and $\sim$7 arcmin$^2$ UVUDF (Teplitz et al.\ 2013) data,
we use a total search area of $\sim$150 arcmin$^2$ to construct
samples of $z\sim2$-3 galaxies.

By combining these data with the optical and near-IR data over the
GOODS North and South fields, we are able to construct a sample of
12098 galaxies in the redshift range $z\sim2$-3.  This is
$>$10$\times$ larger than the samples of $z\sim2$-3 galaxies that
Hathi et al.\ (2010), Oesch et al.\ (2010), and Mehta et al.\ (2017)
had available with {\it HST} in earlier determinations of the $UV$ LF
at $z\sim2$-3.

For our $z\sim4$-9 selections, the most noteworthy new data are the
{\it HST} optical and near-IR observations obtained over the six
parallel fields from the Hubble Frontier Fields program (Lotz et
al.\ 2017).  Those observations probe galaxies to $UV$ luminosities of
$\sim$0.08 $L_{z=3}^{*}$ and are only exceeded by the HUDF in terms of
their sensitivity.  From the six parallel fields to the HFF clusters,
we identify 1381 $z\sim4$, 448 $z\sim5$, 209 $z\sim6$, 122 $z\sim7$,
34 $z\sim8$, and 13 $z\sim9$ galaxies, respectively.  Combining these
samples with those from Bouwens et al.\ (2015), there are $>$12,000
sources in our {\it HST}-based field samples.

All together and including the $z\sim10$-11 selection from Oesch et
al.\ (2018a), our selections of $z=2$-11 galaxies from {\it HST}
fields include 24741 galaxies.  This is more than twice the number of
sources as the largest previous samples of galaxies over this redshift
range.

We leverage the present, even larger samples of $z=2$-9 galaxies to
construct new and improved determinations of the $UV$ LFs at
$z\sim2$-9.  The present determinations constitute the best
blank-field LF results to date.  Encouragingly enough, our new
determinations are in excellent agreement with most previous
determinations where they overlap.

Combining new LF results with the $z\sim10$ LF result from Oesch et
al.\ (2018a), we are in position to reassess the evolution derived in
a self-consistent way, particularly in terms of known Schechter
function parameters like the faint-end slope $\alpha$ and the
normalization $\phi^*$ of the LF.

As in previous studies, we find that the faint-end slope $\alpha$
steepens towards high redshift at approximately a fixed rate
vs. redshift (e.g., Bouwens et al.\ 2015; Parsa et al.\ 2016;
Finkelstein 2016).  The observed evolution appears to be almost
identical to what would expect, i.e., $d\alpha / dz \sim -0.12$ based
on changes to the slope of the halo mass function across the observed
redshift range (e.g., Bouwens et al.\ 2015).

We find that the characteristic luminosity $M^*$ remains relatively
fixed at $\sim-21.02$ mag over the redshift range $z\sim8$ to $z\sim3$
(see also e.g., Bouwens et al.\ 2015; Bowler et al.\ 2015; Finkelstein
et al.\ 2015), but becomes increasingly fainter at $z\lesssim2.5$
where quenching becomes important (e.g., Scannapieco et al.\ 2005;
Peng et al.\ 2010).  The presence of an exponential cut-off in the
$UV$ LF at $z\gtrsim3$ is thought to be imposed by the presence of
dust extinction (e.g., Bouwens et al.\ 2009; Reddy et al.\ 2010), with
the characteristic luminosity set by the $UV$ luminosities where the
increased dust extinction in galaxies more than offsets increases in
the SFRs in galaxies.  The absence of strong evolution in $M^*$
suggests a similar lack of evolution in this transition SFR or $UV$
luminosity.

Finally, we find a systematic decrease in the normalization $\phi^*$
of the $UV$ LF towards high redshift (e.g., McLure et al.\ 2010;
Bouwens et al.\ 2015).  $\log_{10}$ $\phi^*$ can be well described by
quadratic relationship in redshift, with a significantly flatter
relationship at $z<7$ than it is at $z>7$, consistent with the
conclusions from studies favoring ``accelerated'' evolution at $z>8$
(Oesch et al.\ 2012, 2018a).  Interestingly, using the constant star
formation efficiency conditional luminosity function model from
Bouwens et al.\ (2015: their Appendix I), we find that we can
reproduce the observed evolution in $\phi^*$ remarkably well (as shown
in Figure~\ref{fig:modelfit}).  Similar to our discussion in Bouwens
et al. (2015), consistency of the $UV$ LF results with fixed star
formation efficiency models has also been argued in Oesch et
al.\ (2018) and Tacchella et al.\ (2018).  Again, this demonstrates
that much of the evolution in the $UV$ LF (from $z\sim10$ to
$z\sim2.5$ at least) can be explained by the evolution in the halo
mass function and a simple fixed star formation efficiency model.

In a follow-up paper, we will be revisiting the present constraints on
the $UV$ LF at $z\sim2$-9 by incorporating constraints available from
the HFF lensing cluster observations.  With lensing cluster
observations, we will show that we can obtain a completely consistent
constraints on the Schechter parameters using the lensing cluster data
alone or in combination with the field constraints.

\acknowledgements

We acknowledge the support of NASA grants HST-AR-13252, HST-GO-13872,
HST-GO-13792, and NWO grants 600.065.140.11N211 (vrij competitie) and
TOP grant TOP1.16.057.  PAO acknowledge support from the Swiss
National Science Foundation through the SNSF Professorship grant
190079 'Galaxy Build-up at Cosmic Dawn'. The Cosmic Dawn Center (DAWN)
is funded by the Danish National Research Foundation under grant
No. 140.  This paper utilizes observations obtained with the NASA/ESA
Hubble Space Telescope, retrieved from the Mikulski Archive for Space
Telescopes (MAST) at the Space Telescope Science Institute
(STScI). STScI is operated by the Association of Universities for
Research in Astronomy, Inc. under NASA contract NAS 5-26555.

\end{document}